\documentclass[twocolumn,showpacs,nofootinbib,preprintnumbers,amsmath,amssymb,
showkeys,superscriptaddress]{revtex4-1}

\usepackage[pdftex]{graphicx}  
\usepackage{color}
\usepackage{slashed}
\usepackage[normalem]{ulem}
\usepackage{soul}
\usepackage{dsfont}
\usepackage{hyperref}
\usepackage[capitalise]{cleveref}
\usepackage[font=small,format=plain]{caption}
\usepackage{subcaption}

\newcommand \beq{\begin{equation}}
\newcommand \eeq{\end{equation}}

\begin{document}

\title{Two-loop three-gluon vertex from the Curci-Ferrari model\\ and its leading infrared behavior to all loop orders}

\author{Nahuel Barrios\vspace{.4cm}}%
\affiliation{%
Instituto de F\'{\i}sica, Facultad de Ingenier\'{\i}a, Universidad de
la Rep\'ublica, J. H. y Reissig 565, 11000 Montevideo, Uruguay.
\vspace{.1cm}}%
\affiliation{%
Centre de Physique Th\'eorique (CPHT), CNRS, Ecole Polytechnique,\\ Institut Polytechnique de Paris,  Route de Saclay, F-91128 Palaiseau, France.
\vspace{.1cm}}%

\author{Marcela Pel\'aez\vspace{.4cm}}%
\affiliation{%
Instituto de F\'{\i}sica, Facultad de Ingenier\'{\i}a, Universidad de
la Rep\'ublica, J. H. y Reissig 565, 11000 Montevideo, Uruguay.
\vspace{.1cm}}%

\author{Urko Reinosa}%
\affiliation{%
Centre de Physique Th\'eorique (CPHT), CNRS, Ecole Polytechnique,\\ Institut Polytechnique de Paris,  Route de Saclay, F-91128 Palaiseau, France.
\vspace{.1cm}}%

\date{\today}

\begin{abstract}
We evaluate the three-gluon vertex with one vanishing external momentum within the Curci-Ferrari (CF) model at two-loop order and compare our results to Landau-gauge lattice simulations of the same vertex function for the SU(2) and SU(3) gauge groups in four dimensions. This extends previous works \cite{gracey:2019,barrios:2020} that considered similarly the two-loop ghost and gluon two-point functions as well as the two-loop ghost-antighost-gluon vertex (with vanishing gluon momentum). The parameters of the model being adjusted by fitting the two-point functions to available lattice data, our evaluation of the three-gluon vertex arises as a pure prediction. We find that two-loop corrections systematically improve the agreement between the model and the lattice data as compared to earlier one-loop calculations, with a better agreement in the SU(3) case as already seen in previous studies. We also analyze the renormalization scheme dependence of our calculation. In all cases, this dependence diminishes when two-loop corrections are included, which is consistent with the perturbative CF paradigm. In addition, we study the low momentum regime of the three-gluon vertex in relation with the possibility of zero-crossing. Within the CF model, we show that the leading infrared behavior of the exact vertex is given by the same linear logarithm that arises at one-loop order, multiplied by the all orders cubic ghost dressing function at zero-momentum (we provide similar exact results for other vertex functions). We argue that this property remains true within the FP framework under the assumption that the resummed gluon propagator features a decoupling behavior. This shows that the zero-crossing is a property of the exact three-gluon vertex function. Within the CF model, we find however that the scale of the zero-crossing is considerably reduced when going from one- to two-loop order. This seems consistent with some recent lattice simulations \cite{catumba:2022}. Finally, our analysis also allows us to support recent claims about the dominance of the tree-level tensor component \cite{Pinto-Gomez:2022brg}.
\end{abstract}


\pacs{}
\keywords{}
\maketitle

\section{Introduction}\label{sec:intro}

A complete picture of the infrared (IR) regime of Quantum Chromodynamics (QCD) remains still elusive, in particular, in regard to the two phenomena of singular importance for the strong interaction: the confinement of colored asymptotic states as well as the dynamics driving the spontaneous breaking of chiral symmetry and the associated generation of mass.\footnote{A related open question is that of the precise structure of the QCD phase diagram, with new possible phases being recently proposed \cite{McLerran:2007qj,Kojo:2009ha,Glozman:2017dfd,Glozman:2022lda}.} The main difficulty lies in that these phenomena occur at scales such that the QCD coupling constant is not small and standard perturbative methods do not apply. To cope with this, several non-perturbative approaches have been developed over the last decades. The most prominent of them is certainly lattice QCD, a fully non-perturbative, first-principle approach which has allowed for the calculation of many hadronic properties \cite{alexandrou:2008,fodor:2012,carrasco:2014,bazavov:2018}. However, owing to both the enormous computational resources that Monte-Carlo simulations require, and the intrinsic uncertainty that lattice simulations have in the continuum limit, various approaches have been devised directly in the continuum, see below.

As opposed to the lattice set-up, however, any continuum approach requires working within a given gauge and, very often, the Landau gauge is chosen due to its simplifying features. The by now standard gauge fixing procedure is based on the Faddeev-Popov (FP) construction \cite{faddeev:1967} that requires introducing auxiliary fields such as the ghost and antighost fields. This procedure is not exact in the Landau gauge, however, for it disregards the existence of Gribov copies \cite{gribov:1978}. Nevertheless, the FP procedure is expected to become rigorous in the ultraviolet (UV) regime where asymptotic freedom \cite{politzer:1973,gross:1973} allows one to test it by means of a perturbative analysis. In contrast, there is no good reason to believe that the FP procedure extends without modification to the IR regime. As a matter of fact, as one insists in performing perturbative calculations within the FP framework at lower and lower scales, one eventually hits a Landau pole. The latter is usually attributed to an abusive use of perturbative methods beyond their range of validity. However, it could also very well be that, in the Landau gauge, the Landau pole signals, instead, an abusive use of the FP gauge-fixing procedure itself. This dichotomy of interpretation is reflected into the various approaches that are being currently followed to tackle QCD/YM theories in the continuum and which can be classified in two main categories.

The first category is composed of non-perturbative continuum approaches which are usually gathered under the name of functional methods and include for instance the Dyson-Schwinger equations (DSE) \cite{smekal:1997,atkinson:1998,alkofer:2001,fischer:2002,lerche:2002,zwanziger:2002,maas:2004,aguilar:2008,aguilar:2008b,boucaud:2008,boucaud:2008b,huber:2008,dalli:2012,huber:2013,huber:2016,huber:2020,huber:2020b,fischer:2020}, the functional renormalization group (FRG) \cite{fischer:2009,cyrol:2016,ellwanger:1996,pawlowski:2004,fischer:2004,fischer:2007,dupuis:2021}, the Hamiltonian formalism \cite{schleifenbaum:2006,quandt:2014,quandt:2015} and NPI effective action frameworks. In general, such approaches promote the FP action to the non-perturbative level by reformulating the corresponding theory exactly in the form of a system of infinite coupled equations, which must be solved in order to determine the correlation functions for the primary fields, from which one eventually aims at reconstructing the relevant observables. In practice, however, the only way to accomplish this goal is by truncating the infinite tower of equations. It is thus a central question which truncations lead to an accurate description of the correlation functions and thus to an accurate description of the observables. 

In this respect, an important source of complementary information comes from (gauge-fixed) lattice simulations that developed parallel to functional methods both in pure Yang-Mills (YM) theory \cite{mandula:1987,bonnet:2000,cucchieri:2008,duarte:2016,maas:2020,aguilar:2021,catumba:2022} and in QCD \cite{bowman:2004,silva:2011,kizilersue:2021}, mostly, but not only, in the Landau gauge. In particular, one of the central results of the lattice simulations in this gauge is that the gluon two-point function features what is known as a {\it decoupling behavior,} reaching a finite non-zero value at zero momentum \cite{mendes:2010,bogolubsky:2009,bonnet:2000,bonnet:2001,bornyakov:2010,iritani:2009,maas:2013,oliveira:2012,duarte:2016,cucchieri:2008}. Although of fundamental importance because it radically changes our view on the nature of gluons in the infrared, the origin of this dynamically generated mass is still under debate. It could for instance be related to the Schwinger mechanism \cite{cornwall:1982,aguilar:2019b} or it could stem from dynamically generated condensates \cite{verschelde:2001,browne:2003,dudal:2008,Horak:2022aqx,Gies:2022mar}.

A second category of approaches intend, instead, to identify the gauge-fixed action beyond the FP prescription prior to any particular choice of computational scheme. This can be done either by following a semi-constructive approach where one tries to incorporate, at least partially, the effect of the Gribov copies, or, by following a more phenomenological approach where one proposes new terms to the gauge-fixed action beyond those of the FP procedure and tries to constrain them by comparison to experimental results or lattice data. The main representative of the first type of strategy is the Gribov-Zwanziger formalism \cite{gribov:1978,zwanziger:1989,zwanziger:1993,vandersickel:2012,dudal:2008} where one eliminates the infinitesimal copies. The ultimate goal of this formalism is to reduce the functional integrals to a region of the gauge field configuration space where no Gribov copies are present. Unfortunately, this objective has turned out to be extremely difficult to achieve.  As for the second type of strategy, the main representative is the one based on a massive extension of the FP action in the Landau gauge,\footnote{It is worth mentioning that a gluonic mass operator has also been considered in a series of articles \cite{siringo:2016,siringo:2018,siringo:2019,comitini:2020,comitini:2021}. In contrast to the CF approach, the underlying hypothesis in this case is that the FP action remains unchanged in the IR. The gluonic mass operator enters into the picture by being formally added and subtracted to the FP action. This allows for a reorganization of standard perturbation theory which avoids its bad behavior in the IR.} which corresponds to a particular case of the Curci-Ferrari model \cite{curci:1976} and which we address in this work. 

The rationale for adding a mass term is the above mentioned decoupling behavior of the gluon propagator combined with the idea that this behavior could as well be a consequence of taking into account the Gribov ambiguity \cite{serreau:2012,tissier:2018}.  Although phenomenological in nature,\footnote{Recently, a connection between the CF model and the dynamical generation of condensates within the FP framework has been established \cite{Dudal:2022nnu}.} the CF model benefits from very interesting properties which make its study worthwhile. In particular, some of its renormalization group trajectories are regular at all scales (no Landau pole). Moreover, among these trajectories, there is typically one that allows one to reproduce the lattice data for the YM ghost and gluon propagators to a surprisingly good accuracy already at one-loop order. Finally, the corresponding coupling remains relatively small over all scales,\footnote{More precisely, the perturbative expansion parameter $g^2N/(16\pi^2)$ is found to lie below $1$ both in the SU(2) case and in the SU(3) case.} which explains a posteriori why the one-loop calculations provide already a good account of the two-point functions. 

We stress that this last result, although surprising at first sight is not only a result within the CF model but also a prediction of lattice YM theory \cite{bogolubsky:2009,boucaud:2012,bonnet:2000,bonnet:2001,bornyakov:2010,iritani:2009,maas:2013,oliveira:2012} that the CF model just turns out to reproduce. Also, it is not incompatible with the fact that QCD is strongly coupled because, in the presence of quarks, the quark-gluon interaction is typically two to three times larger than the purely gluonic interaction. What this result tells in essence is that QCD could be more subtly non-perturbative than originally thought. While remaining non-pertubative, it could possess a perturbative (pure glue) sector in the infrared, at least with regard to certain quantities.\footnote{As a model beyond the FP gauge-fixing, the CF model has so far essentially been introduced and tested in the Euclidean. This is because the lattice results it relies upon are Euclidean results. It is not clear, to date, whether the CF model relates, through a Wick rotation, to its Minkowskian counterpart.} These ideas have been tested beyond the calculation of vacuum correlators. A recent review on the CF model and its applications to infrared YM/QCD can be found in \cite{Pelaez:2021tpq}. In particular, it has been shown that the perturbative CF model gives a good account of the phase structure of both YM theory \cite{reinosa:2016,reinosa:2015,reinosa:2015b,reinosa:2017,reinosa:2015c,egmond:2022} and QCD in the limit where all quarks are considered heavy \cite{maelger:2018,maelger:2018b}. In the case of QCD, it has recently led to the proposal of a CF-based approach in the form of a double expansion in terms of the pure gauge vertices as well as the inverse number of colors \cite{pelaez:2017,Pelaez:2020ups}. The benefit of this expansion scheme is that the error is in principle controlled by two small parameters.

With the purpose of further testing the soundness of the perturbative CF paradigm with regard to the YM sector, several correlation functions have been evaluated by means of perturbation theory and the results compared to available YM Monte-Carlo simulations. Beyond the original one-loop calculations of the ghost and gluon two-point functions \cite{tissier:2010,tissier:2011}, the ghost-antighost-gluon and three-gluon vertices have been evaluated for all configuration of momenta \cite{pelaez:2013} showing a rather good agreement with the lattice results with a maximal estimated error of around $10-20\%$. Recently, the calculation of two-point functions has been extended to two-loops \cite{gracey:2019}, with a maximal error of around $5-10\%$. These results confirm that the CF model is able to reproduce two-point functions to a high level of accuracy and that the perturbative expansion is under control.\footnote{For similar calculations including quarks, see Refs.~\cite{pelaez:2014,pelaez:2015,pelaez:2017,barrios:2021,figueroa:2022}.} Similar results have been obtained for the ghost-gluon vertex in the limit of vanishing gluon momentum \cite{barrios:2020} with the added strength that this calculation appears as a pure prediction of the model because the parameters were already determined when fitting the lattice gluon and ghost dressing functions in Ref.~\cite{gracey:2019}. 

In this article, we pursue this type of analysis by studying the two-loop CF prediction for the three-gluon vertex  in the limit where one of the external gluon momenta vanishes. The study of the three-gluon vertex is of importance in order to clarify whether or not it displays a zero crossing in the IR. This work comes to complement several calculations that, on the last years, have permitted a better understanding of this fundamental quantity on the lattice \cite{cucchieri:2006,cucchieri:2008,athenodorou:2016,duarte:2016b,vujinovic:2019} and within functional methods \cite{aguilar:2021,blum:2014,eichmann:2014,blum:2015,cyrol:2016,boucaud:2017,aguilar:2020,corell:2018,aguilar:2019}.

The paper is organized as follows. In the next section, we briefly introduce the CF model as well as the renormalization scheme that we use throughout our analysis. Section \ref{sec:three} is devoted to discussing general properties of the three-gluon vertex, in particular its asymptotic UV and IR behaviors. In the IR, we provide an exact formula for the dominant asymptotic contribution, which is essentially that of the one-loop result decorated by a factor that takes the form of the cube of the exact ghost dressing function at zero-momentum. This formula unambiguously shows that zero-crossing has to occur within the CF model, although it does not provide an exact formula for the scale of the zero-crossing. We also argue that the exact formula for the leading IR behavior extends to the case of the FP model within the additional assumption that the gluon propagator resums into a decoupling type propagator. Our analysis corroborates and complements the results of Refs.~\cite{Aguilar:2013vaa,pelaez:2013} from a different perspective. After these generalities, we provide details on the evaluation of the three-gluon vertex at two-loop order in Sec.~\ref{sec:two} and we consider a certain number of crosschecks in Sec.~\ref{sec:check}, including the above mentioned formula for the leading IR asymptotic behavior. In Sec.~\ref{sec:results}, we present the comparison of our results to various lattice data in the SU(2) and SU(3) cases. After some concluding remarks, App.~\ref{ap:diagrams} lists all the two-loop diagrams that were evaluated in this work while Apps.~\ref{app:AI} and \ref{app:AI_CF} present a detailed analysis of the IR structure of the CF model, based on the notion of asymptotically irreducible diagrams \cite{Smirnov:2002pj}. This analysis is the one that eventually leads to the exact formula for the leading IR asymptotic behavior of the three-gluon vertex. Similar exact formulas apply to the four-gluon vertex, as well as to the leading irregular (in the sense of non-Taylor) part of the two-point function.

The article has various entry points depending on the background/interests of the reader. Readers interested in the comparison to the lattice data can read Sec.~\ref{sec:CF} and the beginning of Sec.~\ref{sec:three} where the computed quantities are defined, and then jump directly to Sec.~\ref{sec:results}. Readers interested in the details of the two-loop calculation can read Secs.~\ref{sec:CF} and \ref{sec:three} up to and including \ref{sec:RG} and then dive into Secs.~\ref{sec:two} and \ref{sec:check}, together with App.~\ref{ap:diagrams}. Finally, readers whose interest is in the structural aspects of the CF model in the infrared can read Secs.~\ref{sec:IR} and \ref{sec:check_IR}, supplemented with Apps.~\ref{app:AI} and \ref{app:AI_CF}.

\section{The Curci-Ferrari model}\label{sec:CF}
The Euclidean Curci-Ferrari model is defined in terms of the following Lagrangian density:
\beq
\mathcal{L}=\frac{1}{4}(F_{\mu \nu}^a)^2+\partial_\mu \bar{c}^a(D_\mu c)^a+ih^a \partial_\mu A^a_\mu+\frac{m_B^2}{2}(A_\mu^a)^2\,,\label{eq:cf-model}
\eeq 
where Latin indices denote the generators of the SU(N)  color group. The covariant derivative in the adjoint representation reads
\beq
(D_\mu c)^a\equiv \partial_\mu c^a+g_B f^{abc}A_\mu^b c^c\,,
\eeq
where $f^{abc}$ are the structure constants of the gauge group and $g_B$ the bare gauge coupling constant. The field-strength tensor is defined as
\beq
F_{\mu \nu}^a\equiv \partial_\mu A_\nu^a-\partial_\nu A_\mu^a+g_B f^{abc}A_\mu^b A_\nu^c\,.
\eeq
The model is regularized by means of dimensional regularization in $\smash{d=4-2\epsilon}$ dimensions in which case $g_B$ acquires dimension $\epsilon$. As usual, bare and renormalized fields relate through the renormalization factors $Z$:
\beq
A_B^{a,\mu}=\sqrt{Z_A}A^{a,\mu},\quad c_B^a=\sqrt{Z_c}c^a,\quad \bar{c}^a_B \sqrt{Z_c}\bar{c}^a\,,
\eeq
in the same manner as the bare parameters
\beq\label{eq:rescaling}
\mu^{-\epsilon}g_B=Z_g g,\quad m^2_B=Z_{m^2}m^2\,,
\eeq
where the subscript ``B" refers to bare quantities and $g$ is the (dimensionless) renormalized coupling. For convenience, we introduce $\smash{\lambda\equiv g^2N/(16\pi^2)}$.

The CF model benefits from certain renormalization schemes which are Landau pole free \cite{wschebor:2008}. One such scheme is known as the infrared safe renormalization scheme (IRS) \cite{tissier:2011}, defined by the following renormalization conditions\footnote{See Ref.~\cite{DallOlio:2020xpu} for other Landau pole free schemes within the CF model.}
\beq\label{eq:non-ren-theo}
Z_g \sqrt{Z_A}Z_c=1\,, \quad Z_{m^2}Z_A Z_c=1
\eeq 
and\footnote{For convenience, we have chosen the scale that enters the renormalization conditions equal to the scale that allows one to make the renormalized coupling dimensionless, see Eq.~(\ref{eq:rescaling}). This is not mandatory, however, for these two scales are different in nature, the second one being a regulating scale rather than a renormalization scale. In some instances, it can be useful to take these two scales different of each other \cite{barrios:2021}.}
\beq
G^{-1}(p=\mu)=\mu^2+m^2(\mu)\,,\quad D^{-1}(p=\mu)=\mu^2\,,\label{is-cond-prop}
\eeq 
where $G$ and $D$ refer to the gluon and ghost propagators, respectively, and Eq.~(\ref{eq:non-ren-theo}) is an extension of two non-renormalization theorems \cite{wschebor:2008} in which the finite parts are included.

At one- and two-loop order within the IRS scheme, the CF model features two classes of renormalization group trajectories in the space of dimensionless parameters $(m^2/\mu^2,g^2)$ \cite{reinosa:2017,gracey:2019}. For trajectories in the first class, the flow becomes singular at a finite scale or Landau pole. As for the other trajectories, they are well defined for the entirety  of the renormalization scale range and are characterized by a bounded coupling which approaches zero both in the UV limit and in the IR limit. In principle, in order to work within the perturbative approach, only the latter class of trajectories is rigorously valid. Interestingly enough, these are the flows which best describe YM correlation functions on the lattice \cite{tissier:2011,reinosa:2017}.

\section{The three-gluon vertex}\label{sec:three}
The color structure of the bare three-gluon vertex is given by \cite{Smolyakov:1980wq}
\beq
\Gamma^{(3)B}_{A_\mu^a A_\nu^b A_\rho^c}(p,k,r)=-ig_B f^{abc}\Gamma_{\mu \nu \rho}^B (p,k,r)\,.
\eeq
We shall consider the particular kinematical configuration where the momentum of one of the external gluons vanishes, also known as {\it asymmetric configuration:}
\beq \label{def:3-gluon-vertex}
\Gamma_{\mu \nu \rho}^B(p,-p,0)\equiv \Gamma_{\mu \nu \rho}^B(p)\,.
\eeq
Its corresponding renormalized expression 
\beq
\Gamma_{\mu\nu\rho}(p,\mu)\equiv Z_A^{3/2}Z_g\Gamma_{\mu \nu \rho}^B(p)\,,
\eeq 
admits the following tensorial decomposition\footnote{It is implicitly understood here that renormalized quantities depend on the renormalization scale. We shall make this dependence explicit when discussing the renormalization group in the next section.}
\beq \label{def:3-gluon-tens}
\Gamma_{\mu \nu \rho}(p,\mu)\!=\!2\Gamma_{a}(p^2,\mu)\delta_{\mu \nu}p_\rho+\Gamma_{b}(p^2,\mu)(\delta_{\mu \rho}p_\nu+\delta_{\nu \rho}p_\mu)\,.
\eeq
We are interested in comparing our results with lattice data from Refs.~\cite{aguilar:2021,catumba:2022} in the SU(3) case and from Ref.~\cite{maas:2020} in the SU(2) case. In all these studies, the renormalized quantity the authors compute is, up to a constant factor (see \cref{sec:results}),
\beq \label{def:3-gluon-maas}
\Gamma(p^2,\mu)=\frac{\Gamma^{\text{tree}}_{\mu' \nu' \rho}(p)P^{\perp}_{\mu' \mu}(p)P^{\perp}_{\nu' \nu}(p)\Gamma_{\mu \nu \rho}(p,\mu)}{\Gamma^{\text{tree}}_{\mu' \nu' \rho}(p)P^{\perp}_{\mu' \mu}(p)P^{\perp}_{\nu' \nu}(p)\Gamma_{\mu \nu \rho}^{\text{tree}}(p)}\,,
\eeq
where $\Gamma_{\mu \nu \rho}^{\text{tree}}(p)$ is obtained from \cref{def:3-gluon-vertex} by switching off the interactions. By inserting \cref{def:3-gluon-tens} into \cref{def:3-gluon-maas}, it is straightforward to see that
\beq
\Gamma(p^2,\mu)=\Gamma_a(p^2,\mu)\,.
\eeq
In what follows, for simplicity, we shall refer to this function as the {\it three-gluon dressing function.}

To gain further insight on $\Gamma(p^2,\mu)$, we now discuss its UV and IR asymptotic behaviors. First, there are some general expectations on these behaviors which will later serve as tests for the two-loop calculation to be presented below. Second, the asymptotic expansions reveal the appearance of logarithms. Some of them are symptomatic of the failure of the perturbative expansion at a fixed renormalization scale and need to be taken care of through renormalization group techniques. Some other logarithms are of physical origin and should therefore be retained in our final result. In particular, they relate to the zero-crossing of the dressing function in the IR.

Readers more interested by the details of the two-loop calculation can jump directly to Sec.~\ref{sec:two} while readers interested in the comparison to the lattice result can jump to Sec.~\ref{sec:results}.

\subsection{UV and Renormalization Group}\label{sec:RG}
Let us start by discussing the UV asymptotic behavior. The superficial degree of divergence of the three-gluon vertex is $\smash{\delta=1}$. Owing to the presence of an extra factor of $p$ between this vertex and the three-gluon dressing function, see Eq.~(\ref{def:3-gluon-tens}), we deduce from Weinberg theorem \cite{weinberg:1960} that $\Gamma(p^2,\mu)$ behaves logarithmically at large momentum.

In general, and as our two-loop calculation will later illustrate, the $n$-loop, order $\lambda(\mu)^n$ contribution contains powers of $\ln p^2/\mu^2$ up to and including $(\ln p^2/\mu^2)^n$. This means that, for a fixed renormalization scale $\mu$, and for large enough $p$ such that $\ln p^2/\mu^2$ becomes of the order of $1/\lambda(\mu)$, all loop orders become of the same order, thus invalidating the use of perturbation theory with a fixed renormalization scale in this range of momenta. To cope with this issue, it is mandatory to work, instead, with a running scale $\mu(p)$ choosen such that $\mu(p) \sim p$ in the UV.

For comparison with the lattice data, however, we need to evaluate the three-gluon dressing function $\Gamma(p^2,\mu_0)$ at a fixed renormalization scale $\mu_0$. The latter can be obtained from the running three-gluon dressing function $\Gamma(p^2,\mu(p))$ by means of the Callan-Symanzik equation \cite{callan:1970,symanzik:1970}. In the case of a purely gluonic vertex function this equation reads
\beq
\left(\mu \partial_\mu  -\frac{1}{2}n_A \gamma_A+\beta_\lambda \partial_\lambda+\beta_{m^2}\partial_{m^2}\right) \Gamma^{(n_A)}=0,
\eeq
where we have defined the $\beta$ functions,
\beq
\beta_X(\lambda,m^2)=\mu \left. \frac{d X}{d\mu}\right|_{\lambda_B,m^2_B}\,,
\eeq
for $\smash{X\in \{\lambda,m^2\}}$, and the anomalous dimensions $\gamma$
\beq
\gamma_Y(\lambda,m^2)=\left. \mu \frac{d \ln Z_Y}{d\mu}\right|_{\lambda_B,m^2_B}\,,
\eeq
for $\smash{Y\in \{A,c\}}$. The solution of the Callan-Symanzik equation can be written formally as
\beq
\begin{split}
& \Gamma^{(n_A)}(p,\mu_0,\lambda_0,m^2_0)\\
& \qquad=z_A(\mu;\mu_0)^{-n_A/2}\Gamma^{(n_A)}(p,\mu,\lambda(\mu),m^2(\mu))\,,\label{eq:rewriting}
\end{split}
\eeq
which relates the vertex functions at two different renormalization scales, with
\beq
\ln z_A(\mu;\mu_0)=\int_{\mu_0}^{\mu}\frac{d\mu'}{\mu'}\gamma_A(\lambda(\mu'),m^2(\mu'))\,.
\eeq
The benefit of the rewriting (\ref{eq:rewriting}) is that the beta functions and the anomalous dimensions can also be safely computed within a perturbative expansion, thus giving access to $\Gamma(p^2,\mu_0)$ from perturbative methods, even in the regime $p\gg\mu_0$. 

By applying the Callan-Symanzik equation to the three-gluon dressing function we find
\beq \label{eq:gamma-cs}
\Gamma(p^2,\mu_0)=\frac{\lambda(\mu)}{\lambda(\mu_0)}\frac{\Gamma(p^2,\mu)}{z_A(\mu,\mu_0)^{3/2}},
\eeq
where $\mu$ will be chosen as $\mu(p)$ with $\mu(p)\sim p$ in the UV. In the IRS-scheme, by using the conditions (\ref{eq:non-ren-theo}) and the fact that the bare parameters do not depend on $\mu$, it is easy to show that
\beq
\gamma_A(\lambda,m^2)=2\frac{\beta_{m 2}}{m^2}-\frac{\beta_\lambda}{\lambda},
\eeq
which leads to
\beq \label{eq:z-A-IS}
z_A(\mu,\mu_0)=\frac{m^4(\mu)}{m^4(\mu_0)}\frac{\lambda(\mu_0)}{\lambda(\mu)}.
\eeq
By inserting \cref{eq:z-A-IS} into \cref{eq:gamma-cs} we arrive finally at
\beq \label{eq:cf-pred}
\Gamma(p^2,\mu_0)=\frac{\lambda^{5/2}(\mu)}{\lambda^{5/2}(\mu_0)}\frac{m^6(\mu_0)}{m^6(\mu)}\Gamma(p^2,\mu)\,.
\eeq

\vglue4mm

\subsection{IR and Zero-crossing}\label{sec:IR}
In the opposite momentum range, we cannot rely on Weinberg theorem. However, the IR structure of the CF model can be analyzed using the notion of asymptotically irreducible subgraphs \cite{Smirnov:2002pj}. This general analysis is presented in Apps.~\ref{app:AI} and \ref{app:AI_CF}. 

In particular, we find that the leading infrared contribution to the bare three-gluon dressing function is associated, at all orders, to an effective ghost loop connecting tree-level ghost-antighost-gluon vertices and bare ghost self-energies $\Sigma_B(k)$, see Fig.~\ref{fig:two}. More precisely, one should retain in the ghost self-energies the leading term in a Taylor expansion of the self-energy at small momentum, $\smash{\Sigma_B(k)\sim \sigma_B k^2}$. By resumming all these contributions, one obtains an expression similar to the genuine one-loop ghost contribution to the three-gluon dressing function with however the difference that each tree-level ghost propagator $1/k^2$ has been replaced by $1/(k^2(1+\sigma_B))$. 

Now, $1/(1+\sigma_B)$ is nothing but the bare ghost dressing function $\smash{F_B(p)\equiv p^2 D_B(p)}$ at zero-momentum. We have thus arrived at the conclusion that the dominating infrared behavior of the bare three-gluon dressing function is generated by the bare one-loop ghost contribution multiplied by the cube of the exact bare ghost dressing function at zero-mometum. Upon renormalization, the bare three-gluon dressing function is multiplied by $Z_A^{3/2}Z_g$ which we rewrite conveniently as 
\beq
(Z_A^{1/2}Z_cZ_g)^3Z_c^{-3}Z_g^{-2}\,.
\eeq
The first factor is finite owing to Taylor's non-renormalization theorem and even equals $1$ in any scheme that involves the first of the two conditions in (\ref{eq:non-ren-theo}). The second factor turns the cube of the exact bare ghost dressing function at zero-momentum into the cube of the exact renormalized ghost dressing function at zero-momentum. Finally, the third factor turns the factor $\lambda_B$ that appears in the evaluation of the one-loop ghost contribution into $\lambda(\mu)$. We have thus arrived at the exact result
\beq\label{eq:exact}
\Gamma(p^2,\mu)\sim \frac{\lambda(\mu)}{24}\ln\frac{p^2}{\mu^2}\times F(0)^3\times (Z_A^{1/2}Z_cZ_g)^3\,,
\eeq
where the last factor would be $1$ in the present scheme and $F(p)$ denotes the exact ghost dressing function $p^2 D(p)$. Finally the factor $1/24$ is the one that arises from the strict one loop calculation \cite{pelaez:2013}. Below, we will explicitly check the validity of this exact formula at two-loop order.

The previous result is important in many respects. First, it shows that zero-crossing is an exact property of the CF model. This conclusion extends to the Faddeev-Popov Landau gauge-fixed theory if we make the additional assumption that the resummed gluon propagator features a decoupling behavior similar to the one observed on the lattice. The previous argument, however, does not provide an exact formula for the scale of the zero-crossing since, the latter depends not only on the pre-factor of the logarithm but also on the scale under the logarithm (which is not determined exactly here) and possibly of higher terms in the infrared expansion. Below, we shall analyze numerically how this scale changes when including the two-loop corrections.\footnote{We mention that, even though this scale is scheme independent and, thus, may look as a characteristic feature of Landau gauge YM theory, it could be sensitive to details of the gauge-fixing in the infrared.}

The second consequence of the previous result is that it shows that the status of the $\ln p^2/\mu^2$ in the IR is rather different from that of similar logarithms in the UV. Indeed, it is well captured by perturbation theory since higher loop order calculations do not increase its power. In contrast, the powers of the logarithms $\ln m^2/\mu^2$ that appear in particular inside $F(0)$ are not constrained and increase with the loop order. Then, choosing the running scale $\mu(p)$ such that $\smash{\mu(p)\sim p}$ in the IR invalidates the use of perturbation theory as $\smash{p\to 0}$. In what follows we shall, consider instead the choice $\smash{\mu(p)=\sqrt{p^2+m_0^2}}$, with $m_0=m(1 \text{ GeV})$.

Before closing this section, let us mention that similar exact results hold for the the two- and four-point gluon vertex functions. For the gluon two-point function, one finds
\begin{eqnarray}
& & \frac{\Gamma^{(2)}(p^2,\mu)-\Gamma^{(2)}(0,\mu)}{p^2}\nonumber\\
& & \hspace{1.0cm}\sim \frac{\lambda(\mu)}{12} \ln\frac{p^2}{\mu^2}\times F(0)^2\times (Z_A^{1/2}Z_c Z_g)^2\,.\label{eq:exact2}
\end{eqnarray}
As for the four-point function, for those tensor components that are singular in the IR, we obtain a formula similar to (\ref{eq:exact}) with $F(0)^4(Z_A^{1/2}Z_cZ_g)^4$ rather than $F(0)^3(Z_A^{1/2}Z_cZ_g)^3$ and a different numerical pre-factor stemming from the corresponding ghost box diagram.

It is also interesting to consider the ghost propagator in similar terms. We have seen in App.~\ref{app:AI_CF} that the leading asymptotic behavior stems from the Taylor expansion of the associated vertex, so it is regular. The next term in the expansion involves the structure in Fig.~\ref{fig:one_loop}. The gluon line needs to be interpreted as a chain of gluon self-energy insertions (more precisely their leading IR piece) connected by massive or massless components of the gluon propagator, see Eq.~(\ref{eq:prop}). In the case, where the connecting lines are all massive, the contribution is regular so it cannot contain any logarithms. On the other hand, when one of the connecting lines is massless and thus of the form $P^\parallel(q)/m^2$ with $\smash{P^\parallel_{\mu\nu}(q)\equiv q_\mu q_\nu/q^2}$, it couples to the longitudinal part of the gluon self-energy insertions and thus all other connecting lines are also massless. These chains of massless gluon lines coupled to longitudinal self energy insertions resum as
\beq
\sum_{n=0}^\infty \frac{P_\parallel(q)}{m_B^2}\left(-\frac{\Pi_{\parallel,B}(0)P_\parallel(q)}{m_B^2}\right)^n=\frac{P_\parallel(q)}{m^2_{\rm B}+\Pi_{\parallel,B}(0)}\,.
\eeq
Similarly, as we have seen above, the ghost self-energy insertions in Fig.~\ref{fig:one_loop} resum into an effective tree-level ghost propagator
\beq
\frac{F_B(0)}{p^2}\,.
\eeq
From this we conclude that potential logarithms at next-to-leading order in $\Gamma^{(2)}_{c\bar c}$ arise, to all orders, from the one-loop contribution multiplied by 
\beq
\frac{m^2_{\rm B}F_B(0)}{m^2_{\rm B}+\Pi_{\parallel,B}(0)}\,.
\eeq
Now, using the non-renormalization theorem for the mass \cite{wschebor:2008,tissier:2011}, we find that this ratio is $1$ and then, the logarithm at next-to-leading order in the IR expansion has a pure one-loop origin. When renormalizing, $\Gamma^{(2)}_{c\bar c}(p)$ gets multiplied by $Z_c$, which combines with a factor $Z_{g^2}/Z_{m^2}$ arising from the factor $\lambda_B$ of the one-loop contribution and the factor $1/m^2_B$ of the next-to-leading order IR expansion, to give
\beq
\frac{Z_{g^2}Z_c}{Z_{m^2}}=\frac{(Z_gZ_cZ_A^{1/2})^2}{Z_{m^2}Z_cZ_A}\,,
\eeq
which is fortunately finite and equal to $1$ in the scheme considered here. All in all, we arrive at the exact prediction for the logarithmic part of the next-to-leading order IR behavior:
\begin{eqnarray}
\Gamma^{(2)}_{c\bar c}(p) & = & p^2(\dots)\nonumber\\
& + & \frac{p^4}{m^2}\frac{(Z_gZ_cZ_A^{1/2})^2}{Z_{m^2}Z_cZ_A}\left(-\frac{\lambda}{4}\ln\frac{p^2}{\mu^2}+\dots\right)+\dots\nonumber\label{eq:gagg}
\end{eqnarray}

One can treat the ghost-antighost-gluon vertex in a similar way, with the difference that there is an extra ghost propagator at one-loop order which brings an extra factor of $F_B(0)$ and thus an extra $Z_c$, there is an extra factor of $g_B$ which brings an extra $Z_g$ and there is an extra factor of $Z_A^{1/2}$ from the external gluon leg. Altogether this implies that, if within a considered tensor component of the vertex there is a logarithm at next-to-leading order of the IR expansion, higher loop corrections do not add new logarithms but modify the prefactor of the one-loop logarithm by the finite factor
\beq
F(0)\times\frac{(Z_gZ_cZ_A^{1/2})^3}{Z_{m^2}Z_cZ_A}\,.
\eeq

Finally, let us note that our analysis also allows us to support recent claims on the dominance of the tree-level tensor component for the three-gluon vertex \cite{Pinto-Gomez:2022brg}. First of all, tree-level dominance is expected within the perturbative CF paradigm (and also within the FP paradigm combined with the assumption of a decoupling type gluon propagator) since corrections to vertex functions beyond their tree-level contribution (when any) are expected to be tiny. This argument is of course too na\"\i ve in the presence of IR singularities such as the ones in the three-gluon vertex. Interestingly enough, however, we can provide an exact formula for these singularities which enables us to investigate their exact tensor structure. In particular, if we consider the unprojected three-gluon vertex for momentum configurations depending on one scale $p>0$, that is $p_i=pa_iu_i$, with $a_i>0$ and $u_i$ a unit vector, it is again possible to argue that the exact leading (logarithmic) contribution in this regime is entirely given by that in the one-ghost-loop diagram multiplied by the cube of the exact ghost dressing function at zero-momentum. Now, it is an easy exercise, see App.~\ref{app:one} to show that the logarithm in $p$ in the one-loop diagram has precisely the structure of the tree-level tensor, thus supporting the observation made in \cite{Pinto-Gomez:2022brg}.

\section{Evaluation at two-loop order}\label{sec:two}
We now proceed to the evaluation of the three-gluon dressing function at two-loop order. We first evaluate its bare counterpart $\Gamma_B(p^2)$ and then proceed with its renormalization.

\subsection{Notation}
At two-loop order, $\Gamma_B(p^2)$ reads
\beq
\Gamma_B(p^2)=1+\lambda_B \Gamma_1(p^2,m_B^2) +\lambda_B^2 \Gamma_2(p^2,m_B^2)\,,
\eeq
where $\Gamma_1(p^2,m_B^2)$ and $\Gamma_2(p^2,m_B^2)$ represent the sum of one- and two-loop diagrams, respectively, after projection along the component $2\delta_{\mu \nu}p_\rho$, and
\beq
\lambda_B\equiv \frac{g_B^2 N}{16 \pi^2}\,.
\eeq
By factoring out $\lambda_B^n$, we make explicit the appropriate power of $g_B^2$ as well as the color factor of each loop contribution. As usual, we have also absorbed a factor $(16\pi^2)^n$ in $\Gamma_n(p^2,m_B^2)$. Since we work in $d=4-2\epsilon$ dimensions, the actual dimension of the coupling, $\mu^\epsilon$, is also absorbed into $\Gamma_n(p^2,m_B^2)$. As a consequence, it is convenient to introduce the following notation in relation to the $d$-dimensional Feynman integrals: 
\beq
\int \frac{d^d q}{(2\pi)^d}\to \int_q\equiv 16 \pi^2 \mu^{2\epsilon}\int \frac{d^d q}{(2\pi)^d}.
\eeq

\subsection{Feynman diagrams}

\begin{figure}[h!]
     \centering
     \begin{subfigure}[b]{0.12\textwidth}
         \centering
         \includegraphics[width=\textwidth]{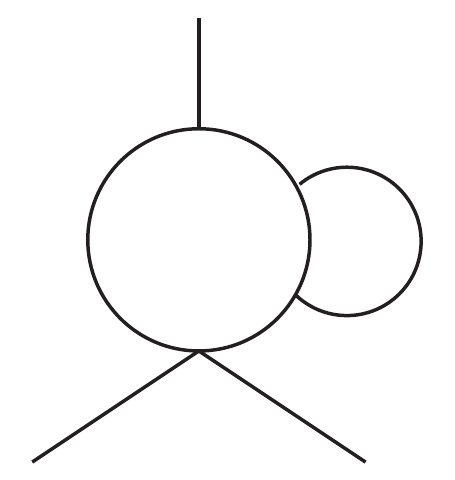}
     \end{subfigure}
          \hfill
     \begin{subfigure}[b]{0.15\textwidth}
         \centering
         \includegraphics[width=\textwidth]{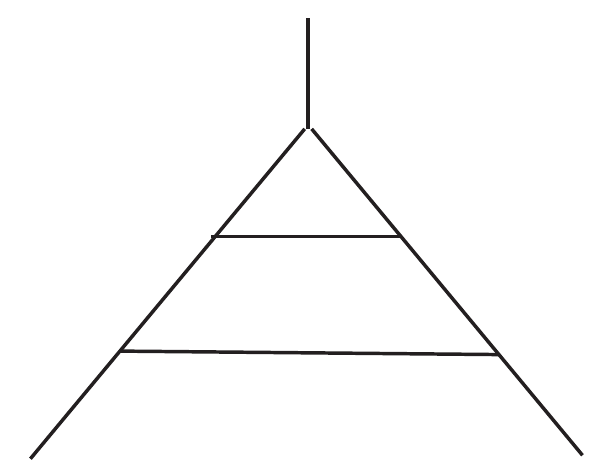}
     \end{subfigure}
     \hfill
     \begin{subfigure}[b]{0.12\textwidth}
         \centering
         \includegraphics[width=.8\textwidth]{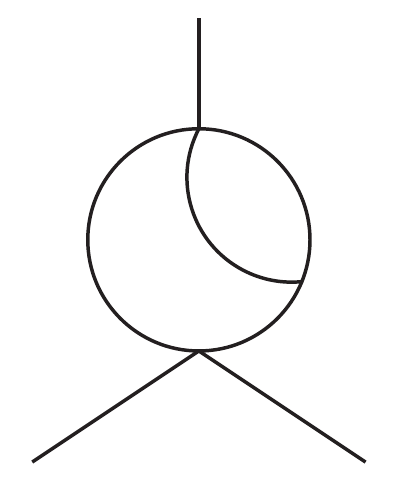}
     \end{subfigure}
        \caption{Examples of topologies of the type $\smash{(V_3,V_4)=(3,1)}$ (left); $(V_3,V_4)=(5,0)$ (middle) and $(V_3,V_4)=(1,2)$ (right).}
        \label{fig:topologies}
\end{figure}

The one-loop diagrams contributing to $\Gamma_1(p^2,m_B^2)$ can be handled essentially by hand, from the writing of the corresponding Feynman integrals, to the evaluation of the latter \cite{pelaez:2013}. As for the  two-loop diagrams contributing to $\Gamma_2(p^2,m_B^2)$, the writing of the Feynman integrals can still be done by hand although we cross-checked the expressions using an automatized routine in \textsc{Mathematica} together with \textsc{qgraf} \cite{Nogueira:1991ex}. For the purpose of organizing the diagrams, it is useful to take into account that the various elements of a Feynman graph in pure YM theory satisfy the relations
\begin{align}
&L=I-(V_1+V_3+V_4-1)\,,\\
&2 I_g+E_g=3 V_3+4V_4+V_1\,,\\
&2 I_{gh}+E_{gh}=2V_1\,,
\end{align}
where $L$ is the total number of loops of the graph, while $I$ and $E$ denote the total number of internal and external lines, respectively. The variables $V_1$, $V_3$ and $V_4$ refer to the number of ghost-antighost-gluon, three-gluon and four-gluon vertices, respectively. Finally, the quantities $I_g$ and $I_{gh}$ correspond to the gluon and ghost internal lines of the graph, respectively.  From these equations, it is easy to show that
\beq \label{eq:diags-topology}
2L+E=V_1+V_3+2V_4+2\,.
\eeq
In the case of the three-gluon vertex, we have $\smash{E=3}$. Morever, since we are interested in two-loop diagrams we set $\smash{L=2}$, which yields 
\beq
5=V_1+V_3+2V_4\,.\label{eq:5}
\eeq
Since the diagrams involving ghosts can easily be deduced from (some of) the diagrams involving only gluons, it is sufficient to list the latter. According to Eq.~(\ref{eq:5}), with $\smash{V_1=0}$, we thus need to consider the three cases $\smash{(V_3,V_4)=(3,1)}$, $(5,0)$ and $(1,2)$, see Fig.~\ref{fig:topologies} for a few examples.

\begin{figure}[h!]
\centering
\raisebox{-0.7cm}{\includegraphics[width=.2\textwidth]{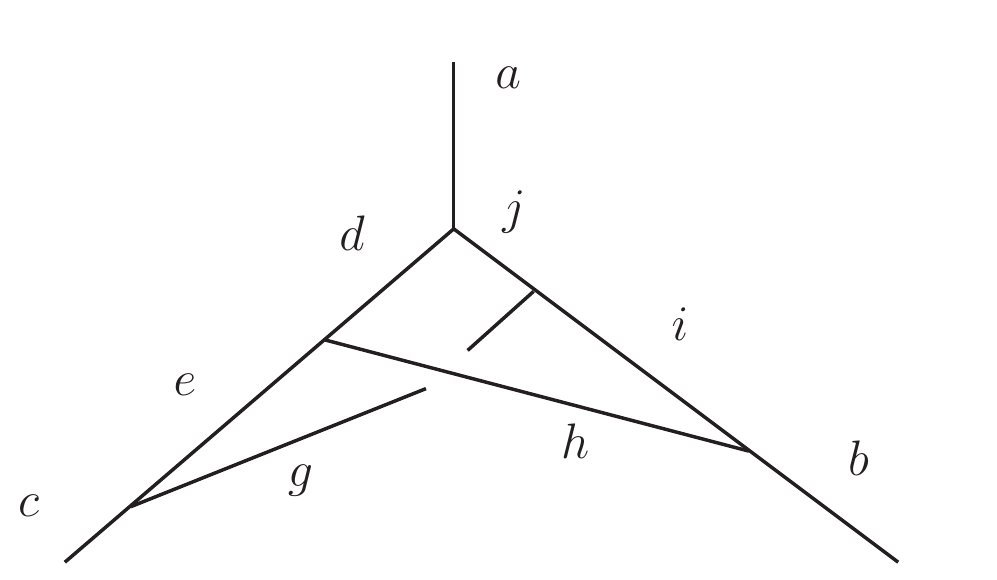}}$\!\begin{aligned}[t]
    &\sim f^{acd}f^{dhe}f^{egc}f^{ibh}f^{jig}\\
   &=\text{Tr}(T^a T^h T^c T^i)f^{ibh}=0,
    \end{aligned}$   
    \caption{The only type of non-planar topology involved in the two-loop evaluation of $\Gamma(p^2)$.}\label{fig:non-planar-diag}
\end{figure}

Non planar diagrams are only of the type  $\smash{(V_3,V_4)=(5,0)}$ and all of them vanish due to their color factor. As shown in \cref{fig:non-planar-diag}, non-planar topologies are proportional to $\text{Tr}(T^a T^h T^c T^i)f^{ibh}$, where the $T^i$ are the generators of the SU(N) gauge group in the adjoint representation. This factor vanishes as it contracts a symmetric tensor with an antisymmetric tensor.  

In total, we evaluated $6$ one-loop diagrams and $72$ two-loop diagrams (excluding non-planar diagrams which vanish anyway). Up to permutations of their external legs, they are drawn in \cref{ap:diagrams}.

\subsection{Reduction to master integrals}
After each diagram contributing to $\Gamma_B(p^2)$ has been written in terms of Feynman integrals, we proceed to the latter in terms of master integrals. As the Feynman diagrams depend on one external momentum only, such reduction leads to self-energy type master integrals. These are:
\begin{align}
A_m & \equiv \int_q G_m(q)\,,\\
B_{m_1 m_2}(p^2) & \equiv \int_q G_{m_1}(q)G_{m_2}(p+q)\,,
\end{align}
at one-loop level, and
\begin{align}
& S_{m_1 m_2 m_3} (p^2) \nonumber \\
& \equiv  \int_q G_{m_1}(q)B_{m_2 m_3}((p+q)^2)\,,\\
& U_{m_1 m_2 m_3 m_4}  (p^2) \nonumber \\
& \equiv  \int_q G_{m_2}(q)G_{m_1}(p+q)B_{m_3 m_4}(q^2)\,,\\
& M_{m_1 m_2 m_3 m_4 m_5}(p^2) \nonumber \\
& \equiv \int_q G_{m_1}(q)G_{m_3}(p+q) \nonumber \\
&\times \int_l G_{m_2}(l)G_{m_4}(p+l)G_{m_5}(l-q)\,,
\end{align}
at two-loop level, with
\beq
G_m(q)\equiv\frac{1}{q^2+m^2}\,.
\eeq
In addition, our results depend on another well-known integral,
\beq
T_{m_1 m_2 m_3} (p^2)=-\frac{\partial S_{m_1 m_2 m_3}}{\partial m_1^2}(p^2)\,.
\eeq
The reduction to master integrals was carried out in \textsc{Mathematica}, via the \textsc{Fire} package \cite{smirnov:2020}, which makes an extensive use of Laporta's algorithm \cite{laporta:2000}. In the case of $\Gamma_1(p^2,m_B^2)$, the output of this procedure is a sum of integrals of the type $A$ and $B$. In a similar manner, $\Gamma_2(p^2,m_B^2)$ is expressed as a sum of integrals of the type $S, U, M, T$ and products of integrals $A$ and $B$. In both cases, the coefficients of the master integrals are given by rational fractions involving $p^2$, $m_B^2$ and the space-time dimension $d$. 

In the reduction of $\Gamma_2(p^2,m_B^2)$ we found some integrals of the type $M$ but with one of the propagators to the power -1. Thankfully, these terms can be rewritten in terms of master integrals. For instance, we found
\beq
\int_l G_0^{-1}(l)G_m(p+l)B_{mm}(l^2)=\frac{p^2-3m^2}{3}S_{mmm}(p^2)+A_m^2.
\eeq
The procedure we followed in order to obtain this type of reduction is the one described in the App. A of Ref.~\cite{barrios:2020}. It can also be crosschecked with \textsc{Fire}.

\subsection{Renormalization}
Up to this point we have been working with the bare three-gluon dressing function. This quantity is UV divergent both at one- and two-loop order, and, with the goal of producing meaningful results it is necessary to proceed with its renormalization. In the present case, this is easily done because the divergent structure of the master integrals (more precisely the coefficients of the corresponding poles in $\epsilon^{-2}$ and $\epsilon^{-1}$) is known analytically \cite{martin:2003}. This divergence are of course absorbed in the renormalization factors that enter the expression of the renormalized three-gluon dressing function:
\beq \label{eq:gamma-ren}
\begin{split}
\Gamma(p^2)=&\sqrt{Z_\lambda} Z_A^{3/2}\left( 1+ \lambda Z_\lambda \Gamma_1(p^2,Z_{m^2}m^2)\right. \\
&\left. +\lambda^2 \ \Gamma_2(p^2,m^2)   \right),
\end{split}
\eeq 
with $\smash{Z_\lambda=Z_{g^2}}$. In the last term, bare quantities can be automatically replaced by renormalized ones, since corrections coming from the renormalization factors contribute to higher orders in the perturbative series. 

In order to arrive at the final renormalized expression at two-loop order, we expand \eqref{eq:gamma-ren} to order $\lambda^2$, neglecting terms of order $\lambda^3$ or higher. To do so, it is important to consider up to two-loop contributions to the renormalization factors,
\begin{align}
Z_X&=1+\lambda Z_{X,1}+\lambda^2  Z_{X,2}+\mathcal{O}(\lambda^3)\,,
\end{align}
where
\begin{align}\label{eq:Z-fact-exp}
Z_{X,1}&=\frac{Z_{X,11}}{\epsilon}+Z_{X,10}+\epsilon Z_{X,1-1}+\mathcal{O}(\epsilon^2)\,,\\
Z_{X,2}&=\frac{Z_{X,22}}{\epsilon^2}+\frac{Z_{X,21}}{\epsilon}+Z_{X,20}+\mathcal{O}(\epsilon)\,,
\end{align}
with $X \in \{A,\lambda,m^2\}$. We then find
\beq \label{eq:gamma-ren-2-loop}
\begin{split}
& \Gamma(p^2)=1+\lambda \left(\frac{3}{2}Z_{A,1}+\frac{1}{2}Z_{\lambda,1}+\Gamma_1(p^2,m^2) \right)\\ 
& + \lambda^2 \left( \frac{3}{8}Z_{A,1}^2+\frac{3}{2}Z_{A,2}+\frac{3}{4}Z_{A,1}Z_{\lambda,1}-\frac{Z_{\lambda,1}^2}{8}\right. \\
&\left. +\frac{Z_{\lambda,2}}{2} +\frac{3}{2}Z_{A,1}\Gamma_1(p^2,m^2)+\frac{3}{2}Z_{\lambda,1}\Gamma_1(p^2,m^2)\right. \\
&\left. +m^2 Z_{m^2,1}\frac{\partial \Gamma_1}{\partial m^2}(p^2,m^2)+\Gamma_2(p^2,m^2) \right) + \mathcal{O}(\lambda^3)\,.
\end{split}
\eeq
The derivative $\partial \Gamma_1/\partial m^2$ generates integrals of the type $\partial A_m /\partial m^2$, $\partial B_{m 0}(p^2)/\partial m^2$ and $\partial B_{m m}(p^2)/\partial m^2$. All of them can be expressed in terms of one-loop master integrals by using integration by parts techniques: 
\begin{align}
\frac{\partial A_m}{\partial m^2}&=\left(\frac{d}{2}-1 \right)\frac{A_m}{m^2}\,,\\
\frac{\partial B_{m0}(p^2)}{\partial m^2}&=\frac{1}{p^2+m^2}\left((d-3)B_{m0}(p^2)+\frac{\partial A_m}{\partial m^2} \right)\,,\\
\frac{\partial B_{m m}}{\partial m^2}(p^2)&=\frac{d-2}{2 m^2 (p^2+4 m^2)}A_m +\frac{d-3}{p^2+4m^2}B_{m m}(p^2)\,.
\end{align}

The IRS renormalization factors were already determined in Ref.~\cite{gracey:2019} from the renormalization of the ghost and gluon two-point functions as well as from the two non-renormalization theorems. Consequently, a first check on the calculation of $\Gamma(p^2)$ consists in verifying that the various divergent terms from \cref{eq:gamma-ren-2-loop} cancel with each other, leading to a finite expression. This is a non-trivial check since our expression for $\Gamma_2(p^2,m_B^2)$ has terms of order $\epsilon^{-1}$, $\epsilon^{-2}$ and even $\epsilon^{-3}$. These triple poles are a result of the reduction from Feynman to master integrals, which can generate spurious poles $(4-d)^{-1}$. More precisely, these are\\


\begin{widetext}
\beq \label{eq:term-ep-m1}
\begin{split}
\frac{(4-d)^{-1}}{96} &\left[ \left(\frac{4}{m^2}-\frac{2}{3p^2}+\frac{7p^2}{3m^4}-\frac{p^4}{3m^6}\right)A_m B_{00}+\left(1-\frac{p^2}{m^2}\right)B_{0 0} B_{m 0} +\left(-\frac{13}{3m^2}+\frac{2m^2}{3p^4}-\frac{2}{p^2}+\frac{p^2}{3m^4} \right)I_{m00}  \right.  \\
&\left. +\left(\frac{11}{3m^2}-\frac{2p^2}{3m^4}-\frac{p^4}{3m^6}\right)S_{000}+\left(\frac{1}{3m^2}-\frac{2m^2}{3p^4}+\frac{8}{3p^2}-\frac{8p^2}{3m^4}+\frac{p^4}{3m^6} \right)S_{m00} \right. \\
&\left. +\left(3-\frac{2m^2}{3p^2}+\frac{10p^2}{3m^2}-\frac{p^4}{3m^4} \right)U_{00m0}+\left(\frac{p^2}{m^2}-1\right)U_{0m00} \right],
\end{split}
\eeq
\end{widetext}
where $I_{m_1 m_2 m_3}$ stands for $S_{m_1 m_2 m_3}(p^2=0)$. We have checked that the triple poles cancel among the various terms in this formula. Indeed, this should be the case since there is no other term in \cref{eq:gamma-ren-2-loop} capable of canceling such terms. We then checked that the terms proportional to $\epsilon^{-2}$ and $\epsilon^{-1}$ in $\Gamma(p^2)$ cancel as well, as it should.

\subsection{Finite parts}

After verifying that our expression is UV-finite, we must carefully look up to which order in $\epsilon$ we should expand the various terms appearing in \cref{eq:gamma-ren-2-loop} so as to not to miss any finite contribution. It is clear that the terms $\Gamma_1(p^2,m^2)$ and $Z_{X,1}$ should be expanded to order $\epsilon^1$. This is because in the products $Z_{A,1}^2$, $Z_{A,1}Z_{\lambda,1}$, $Z_{A,1}\Gamma_1$, $Z_{\lambda,1}\Gamma_1$ and $Z_{m^2,1}\frac{\partial \Gamma_1}{\partial m^2}$, all of which contribute to the term proportional to $\lambda^2$, the two terms involved in each product have poles of the form $1/\epsilon$. Hence, terms of order $\epsilon^1$ yield finite quantities. In contrast, $\Gamma_2(p^2,m^2)$ and $Z_{X,2}$ should be regarded only up to order $\epsilon^0$, since no product involving such quantities intervene at order $\lambda^2$. As a result, one could argue that one-loop master integrals, $A$ and $B$, should be expanded to order $\epsilon^1$ which are known analytically, and two-loop master integrals, $S$, $T$, $U$ and $M$ up to order $\epsilon^0$ which can all be evaluated using the \textsc{Tsil} package \cite{martin:2006}.

However, since the term (\ref{eq:term-ep-m1}) introduces an additional $1/\epsilon$ coming from the reduction to master integrals, we find that it is necessary to expand $A_m$, $B_{00}$ and $B_{m0}$ to order $\epsilon^2$ and $I_{m00}$, $S_{000}$ , $S_{m00}$, $U_{00m0}$ and $U_{0m00}$ to order $\epsilon^1$ in order to keep all the finite contributions. This is in general not covered by the \textsc{Tsil} package. Fortunately, all these expansions were already considered in Ref.~\cite{barrios:2020}.

\section{Crosschecks}\label{sec:check}

The calculation of $\Gamma(p^2)$ involves many diagrams, which, after the reduction to master integrals, generate a significant amount of terms. Consequently, the result for $\Gamma(p^2)$ needs to be tested as much as possible. This section briefly describes some of these tests. All of them obey the same logic: some specific feature is satisfied by $\Gamma(p^2)$ but not by the individual terms which make up $\Gamma_1(p^2,m^2)$ and $\Gamma_2(p^2,m^2)$. As a consequence, very specific cancellations among the terms in \cref{eq:gamma-ren-2-loop} must hold so as to produce a $\Gamma(p^2)$ with the right characteristics.

\subsection{UV asymptotic behavior}
We have already seen that the three-gluon dressing function should behave logarithmically at large momenta. In contrast, the individual terms that compose $\Gamma(p^2,\mu)$ after the \textsc{Fire} reduction can grow much faster. In order to check that these larger contributions cancel among each other, we used UV expansions for all the master integrals involved in $\Gamma(p^2,\mu)$. We determined these expansions by using our own implementation of the algorithm described in Ref.~\cite{Davydychev:1993pg}. Furthermore, a numerical test is not possible in the UV region, since our implementation of \textsc{Tsil} does not have a good behavior in that range of momenta. 

At leading order of the $p\to\infty$ expansion, we find\footnote{In the UV, since the perturbative expansion at large $p$ makes sense only in the presence of a running scale $\mu(p)$ such that $\mu(p)\sim p$, we can also expand with respect to $\mu$.}
\begin{eqnarray} \label{eq:gamma-uv-exp}
\Gamma(p^2,\mu) & = &  1+\lambda(\mu)\Bigg[\frac{37}{24}+\frac{17}{12}\ln\left(\frac{p^2}{\mu^2}\right) \Bigg]\nonumber\\
& + & \lambda^2(\mu)\Bigg[\frac{143}{96}\ln\left(\frac{p^2}{\mu^2}\right)-\frac{51}{32} \ln\left(\frac{p^2}{\mu^2}\right)^2\nonumber\\
& & \hspace{1.0cm} +\,\frac{153}{32}+\frac{5}{16} \zeta (3)  \Bigg]+\mathcal{O}\left(\frac{m^2}{p^2}\right).
\end{eqnarray}
As anticipated, the three-gluon dressing function grows logarithmically in the UV and we observe that the power of the logarithms increases with the loop order thus requiring the use of the renormalization group as discussed in Sec.~\ref{sec:RG}.

\subsection{IR asymptotic behavior}\label{sec:check_IR}

As we have already argued, the exact leading asymptotic infrared behavior of the three-gluon dressing function is given by a linear logarithm, which has essentially a one-loop origin, dressed by the cube of the ghost dressing function at zero-momentum. Expanding the exact formula (\ref{eq:exact}) at two-loop order, we find (we also use the first condition in (\ref{eq:non-ren-theo}))
\beq
\Gamma(p^2,\mu)\sim \frac{\lambda(\mu)}{24}\ln\frac{p^2}{\mu^2}\times (1-3\sigma_1)\,.
\eeq
where $\sigma_1$ denotes the $k^2$-coefficient of the one-loop ghost self-energy $\Sigma(k)$ in the $k\to 0$ limit. 

Similarly to the UV case, this particular IR behavior is not necessarily observed in each of the master integrals which make $\Gamma(p^2,\mu)$. With the purpose of obtaining the IR expansion of the three-gluon dressing function we determined the IR expansions for the various master integrals which compose $\Gamma(p^2,\mu)$. We achieved this by implementing the algorithm described in Ref.~\cite{davydychev:1993}, see also App.~E from Ref.~\cite{barrios:2021}. This algorithm cannot be applied in some cases and a more sophisticated strategy, described in Ref.~\cite{berends:1995} was needed.

At first non-trivial order of the $p\to 0$ expansion, we find
\begin{eqnarray} 
& & \Gamma(p^2,\mu)= 1+\frac{\lambda(\mu)}{24} \Bigg[\ln \left(\frac{p^2}{\mu^2}\right)\left(1-3\tau_1\right)+C_1(m^2/\mu^2)\Bigg] \nonumber\\
& & \hspace{0.5cm}+\,\frac{\lambda^2(\mu)}{24} C_2(m^2/\mu^2) +\mathcal{O}\left(\frac{p^2}{m^2}\right),
\label{eq:gamma-ir-exp}
\end{eqnarray}
with 
\begin{eqnarray}
\tau_1 & = & \frac{\lambda(\mu)}{4}\frac{\mu^2}{m^2}\Bigg(\frac{m^4}{\mu^4} + \frac{5}{2}\frac{m^2}{\mu^2} +\ln\frac{\mu^2}{m^2}\nonumber\\
& &  \hspace{1.0cm}-\,\Bigg(1+\frac{m^2}{\mu^2}\Bigg)^3 \ln\left[1 + \frac{\mu^2}{m^2}\right]\Bigg),
\end{eqnarray}
which can be checked to equal $\sigma_1$ in the present scheme. Our two-loop results are thus compatible with the exact formula (\ref{eq:exact}). We have performed a similar two-loop check of (\ref{eq:exact2}) and (\ref{eq:gagg}), as well as the corresponding prediction for the ghost-antighost-gluon vertex using the two-loop results of Ref.~\cite{barrios:2020}.

\subsection{Regularity at $p^2=m^2$}

Our result for $\Gamma(p^2)$ is not regular at $p=0$. This is a genuine singularity associated with the zero-crossing. In addition, some of the individual contributions to our result presented a singularity at $\smash{p^2=m^2}$. There is nothing special about this specific (Euclidean) configuration. As a consequence, another test on the calculation of $\Gamma(p^2)$ consists in verifying that this divergence is spurious, emerging as a result of the particular reduction from Feynman to master integrals. When adding all the contributions, the corresponding residue reads
\beq \label{eq:dm4term}
\begin{split}
&\frac{\lambda^2}{64}\left[\frac{2-d}{d-1}A_m B_{00}(m^2)-\frac{m^2(d-3)}{d-1}B_{00}(m^2) \right.\\
&\left.+\frac{2-d}{d-1}I_{m00}-\frac{(4-d)m^4}{2(d-1)}M_{0000m}(m^2)\frac{3d-8}{d-1}S_{m00}(m^2) \right].
\end{split} 
\eeq
The one- and two-loop master integrals in the above equation are known analytically in $d=4-2\epsilon$ dimensions to order $\epsilon^1$ and $\epsilon^0$, respectively. This allows one to show that \cref{eq:dm4term} vanishes, at least to order $\epsilon^0$. Of course, higher orders in $\epsilon$ are not relevant in our analysis.

\subsection{Zero mass limit}
We can also consider the zero mass limit of our result. This limit is regular for any $p^2>0$ and has already been calculated in Ref.~\cite{davydychev:1998}. This is here a double check since individual terms in the expression of $\Gamma(p^2)$ are not necessarily regular  and cancellations must occur in order to produce the right limit.

For the purpose of computing this limit, we exploited the fact that, on dimensional grounds, any of the master integrals involved in the reduction of $\Gamma(p^2)$ can be written as
\beq
(\mu^{2\epsilon})^L F(p^2,m^2)=(\mu^{2\epsilon})^L (m^2)^{D/2}F(p^2/m^2,1),
\eeq
where $L$ is the number of loops and $D$ denotes the mass dimension of the integral. From this way of writing the integrals it is clear that the low mass expansion of any master integral is equivalent to the large momentum expansion. As a consequence, $\Gamma(p^2,m^2 \to 0)$ is simply the leading term in the UV expansion given in \cref{eq:gamma-uv-exp} below. Since a different renormalization scheme was used in Ref.~\cite{davydychev:1998}, we resorted to a comparison of the bare results, which do indeed coincide in the Landau gauge.\\

\section{Results}\label{sec:results}

We finally present our results for the two-loop three-gluon dressing function in the CF model and the comparison to SU(2) and SU(3) lattice data. As already mentioned above, the parameters of the model, $m_0$ and $\lambda_0$, are determined by fitting the CF expressions for the gluon and ghost two-point functions to the lattice data. This adjustment of the parameters has already been done in earlier works in the IRS scheme with the choice of running scale $\smash{\mu(p)=p}$. 

As we have argued, however, a more sensible choice is $\smash{\mu(p)=\sqrt{p^2+m_0^2}}$. Thus, for consistency and before investigating the three-gluon dressing function, we redo the fits of the two-point functions with this new choice of running scale. To this purpose, we minimize a joint error $\chi_{DF}$, defined as
\beq
\chi_{DF}^2=\frac{1}{2}\left(\chi_D^2+\chi_F^2\right),
\eeq
where $\chi_X^2$ is given by
\beq
\sum_{i=1}^N \frac{X_{\text{latt.}}^{-2}(\mu_0)+X_{\text{latt.}}^{-2}(p_i)}{2N}(X_{\text{latt.}}(p_i)-\mathcal{N}_X X_{\text{CF}}(p_i))^2,
\eeq
with $X\in\{D,F\}$, $X_{\text{latt}}(p_i)$ and $X_{\text{CF}}(p_i)$ refer to the value of $X(p_i)$ obtained from the lattice and from the Curci-Ferrari model evaluated at $p_i$, respectively. The normalizations ${\cal N}_X$ are also chosen to minimize the joint error (they can be determined analytically in terms of the lattice and model values) and are needed in order to connect the two-point functions in the IRS scheme and the two-point functions in the scheme used on the lattice (only ratios of a two-point function at two different momentum scales are scheme-independent). We mention that the quality of the obtained fits is very similar to that obtained with the choice $\smash{\mu(p)=p}$

Once the parameters of the model have been fixed, our calculation predicts the renormalized three-gluon dressing function $\Gamma(p^2,\mu_0)$ in the IRS scheme which we rewrite as $\Gamma(p^2)$ for simplicity. Since the scheme used in the lattice is different, we must allow for an overall normalization $\mathcal{N}$.\footnote{At an exact level of treatment, it is possible to relate ${\cal N}$, ${\cal N}_A$ and ${\cal N}_c$ provided one works within a scheme where $Z_A^{1/2}Z_cZ_g=1$ holds. However, this relation does not hold exactly at a finite loop order. Moreover, this requires that the lattice simulations for the vertex and the two-point functions to be exactly consistent. For these reasons, we allowed for a normalization ${\cal N}$ independent from ${\cal N}_A$ and ${\cal N}_c$.} The latter is determined so as to minimize the absolute error between the CF prediction \cref{eq:cf-pred} and the lattice data:
\beq
\chi^2_{\Gamma}=\frac{1}{N}\sum_{i=1}^N \left(\frac{\Gamma_{\text{latt.}}(p_i^2)-\mathcal{N}\ \Gamma(p_i^2)}{\Gamma_{\text{latt.}}(p_N^2)}  \right)^2,
\eeq
where the sum runs over the lattice points. We have here opted for a normalized absolute error rather than a relative error. The reason for this choice is that the relative error does not grasp correctly the differene between the CF predictions and the lattice data in the IR region where lattice data are close to zero. As a result of minimizing with respect to $\mathcal{N}$, we get
\beq
\mathcal{N}=\frac{\sum_{i=1}^N \Gamma_{\text{latt.}}(p_i^2)\Gamma(p_i^2)}{\sum_{i=1}^N \Gamma(p_i^2)^2}.
\eeq

\begin{figure}[t]
\begin{subfigure}{.5\textwidth}
\includegraphics[scale=.55]{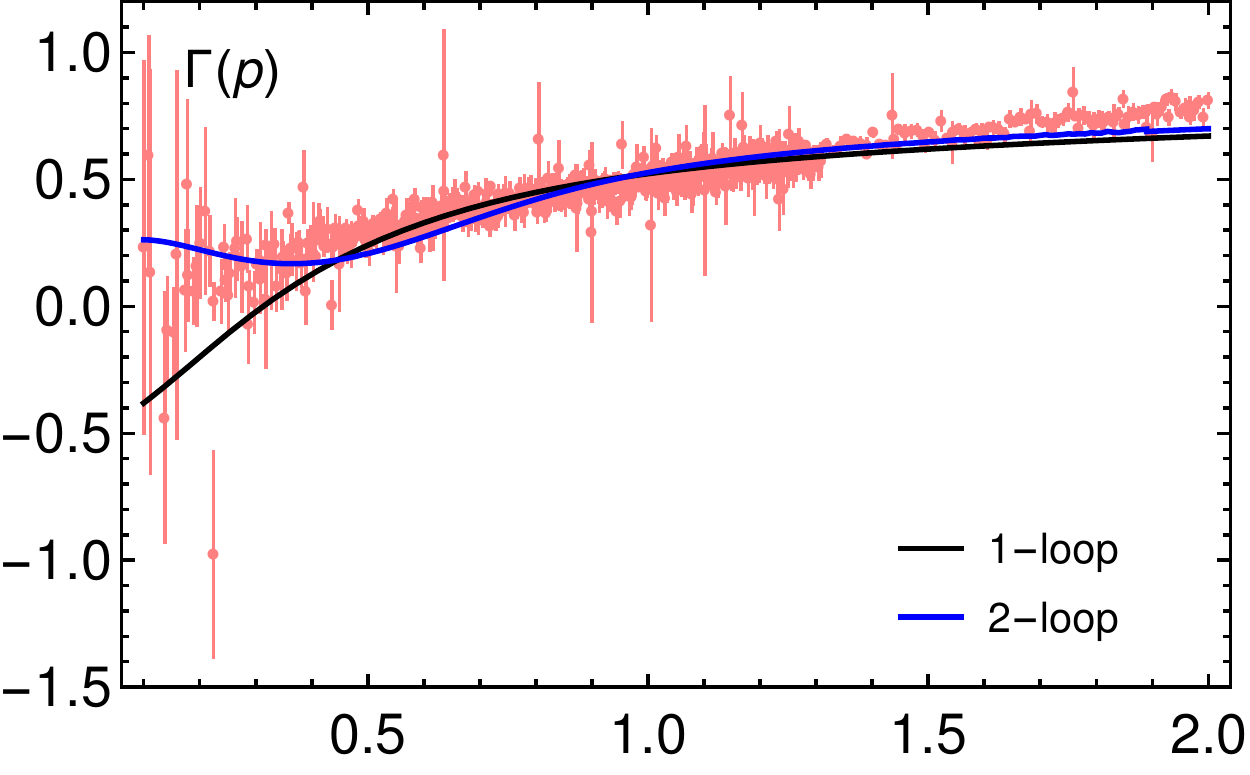}
\end{subfigure} 
\vglue4mm
\begin{subfigure}{.5\textwidth}
\includegraphics[scale=.55]{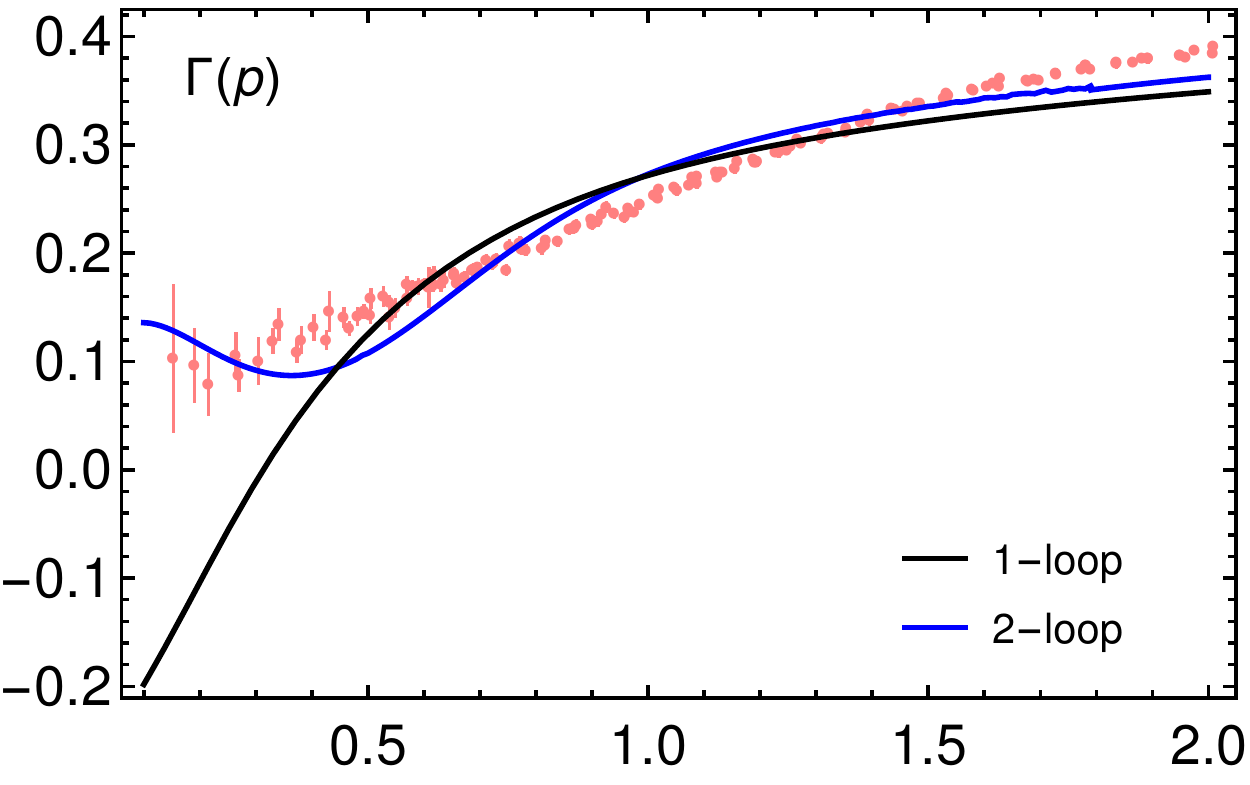} 
\end{subfigure}
\caption{One- and two-loop three-gluon dressing function as predicted within the CF model, compared with the SU(3) lattice results of Ref.~\cite{aguilar:2021} (top) and Ref.~\cite{catumba:2022} (bottom).}\label{fig:SU3}
\end{figure}

\subsection{SU(3)}
In the SU(3) case, we compared our results with two lattice data sets \cite{aguilar:2021,catumba:2022} while the fits of the ghost and gluon two-point functions were done using the lattice data of Ref.~\cite{Duarte:2017wte,Dudal:2018cli}. The comparison between our prediction and the results of Ref.~\cite{aguilar:2021} is displayed in \cref{fig:SU3}. In this plot, as in any other plot from this section, the vertical axis shows the three-gluon dressing function $\Gamma(p)$ while the horizontal axis refers to the momentum in GeV.

As can be seen from the figure, the one- and two-loop predictions are rather consistent with the data. We note that, even though our results feature a zero-crossing, as is evident from  \cref{eq:gamma-ir-exp}, the two-loop corrections move the latter deeper in the infrared, where no lattice data are available. Two-loop corrections also introduce qualitative differences in the IR (as compared to the one-loop results), around 0.5 GeV, where $\Gamma(p^2)$ bends upwards to go downwards again in the deep IR. This is precisely the range in which our coupling is the largest. Of particular interest is the comparison with the data of Ref.~\cite{catumba:2022} where no zero-crossing is observed, at least not within the simulated range of momenta. Although one-loop corrections do display a zero-crossing at a scale of around $350$~MeV, thus contradicting the data, once two-loop corrections, the scale of the zero-crossing is pushed deep in the infrared and the two -loop prediction agrees pretty nicely with the data, see also \cref{tab:error-su3} below.

\begin{table}[t] 
\begin{tabular}{|| c|c|c|c|c|c ||}
\hline
order & $\lambda_0$ & $m_0$ (MeV) & $\chi_{DF}(\%)$ & $\chi_{\Gamma,A}(\%)$  & $\chi_{\Gamma,C}(\%)$ \\
\hline \hline
\;1-loop\; & \;\;0.30\;\; & \;\;350\;\; &\;\;4.6\;\; & \;\;13.0\;\;  & \;\;11.6\;\;   \\
\hline
\;2-loop\; & \;\;0.27\;\;  & \;\;320\;\; &\;\;3.2\;\; & \;\; 10.6\;\; & \;\;5.5\;\;\\
\hline
\end{tabular}
\caption{This table shows, depending on the loop order, the values of the parameters which best fit the  lattice data of Refs.~\cite{Duarte:2017wte,Dudal:2018cli} for the the ghost and gluon two-point functions, the corresponding joint error for the gluon and ghost dressing functions, and the individual errors for the predicted three-gluon dressing function in comparison to the data of Refs.~\cite{aguilar:2021} and \cite{catumba:2022}, denoted respectively $\chi_{\Gamma,A}$ and $\chi_{\Gamma,C}$.}\label{tab:error-su3}
\end{table}

\cref{tab:error-su3} collects the values of the errors $\chi_{DF}$ and $\chi_\Gamma$, at one- and two-loop order and for both data sets. We observe a decrease of all errors when two-loop corrections are included. In the particular case of the lattice data set from  Ref.~\cite{catumba:2022}, as already mentioned above, we find a much smaller error at two-loop order. This could be attributed to a smaller uncertainty of the data in comparison to the lattice results of Ref.~\cite{aguilar:2021}, particularly in the deep IR, where the data display larger errors. In any case, both the one- and two-loop calculations of $\Gamma(p^2)$ are totally compatible with lattice data from \cite{aguilar:2021}. We emphasize, again, that these results are a pure prediction of the CF model. 

As a result, we can conclude that, in line with previous results, the perturbative CF model can describe the results of lattice YM theory both at a qualitative and at a quantitative level. On top of this, successive perturbative orders tend to be more accurate, which is consistent with a controlled perturbative approach.

For completeness, we mention that we have redone the analysis with the choice $\mu(p)=p$ and the differences with the choice $\mu(p)=\sqrt{p^2+m_0^2}$ are numerically small (of around $0.4\%$). The latter choice tends to slightly improve the one-loop fits, while it worsens the two-loop fits by the same amount.

\subsection{SU(2)}
\begin{figure}[t]
\includegraphics[scale=.55]{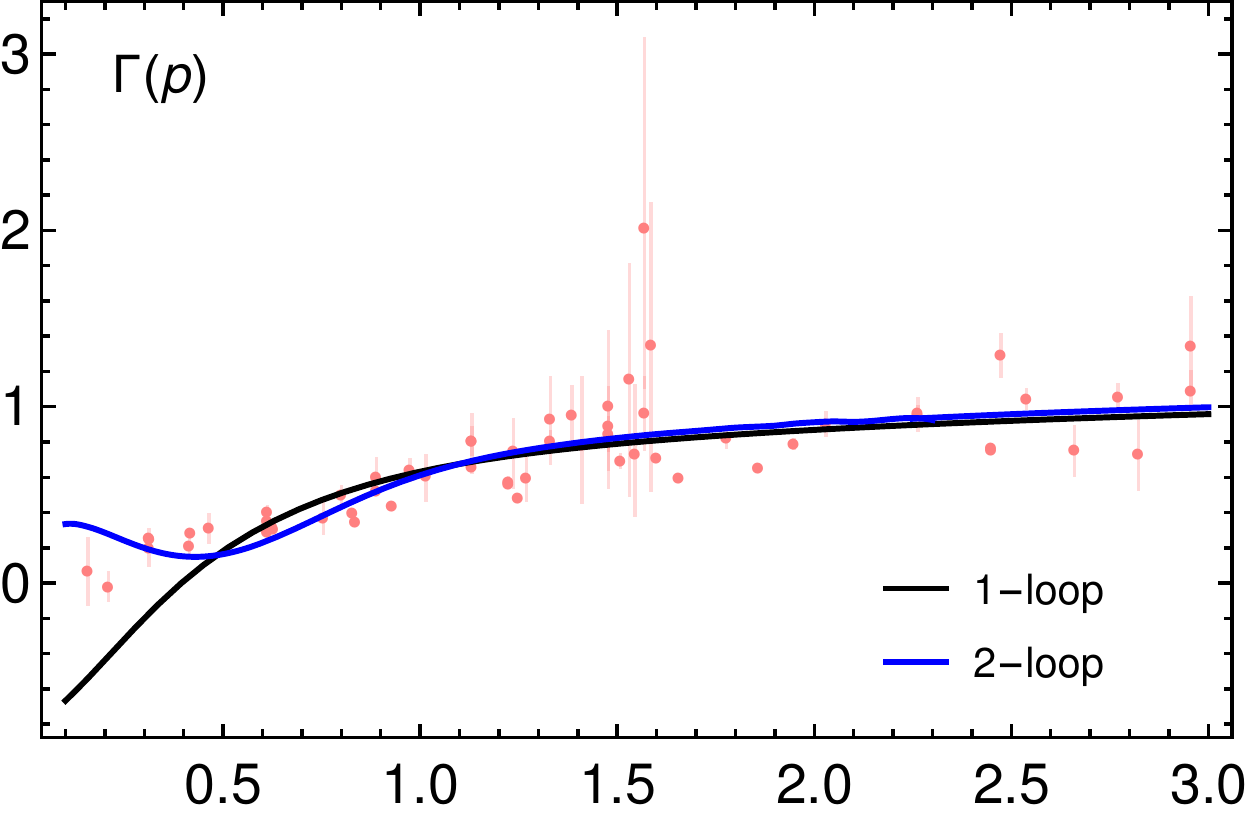} 
\caption{One- and two-loop three-gluon dressing function as predicted within the CF model, compared with the SU(2) lattice results of Ref.~\cite{maas:2020}.}\label{fig:SU2}
\end{figure}

In the SU(2) case, the fits of the gluon and ghost dressing functions were carried out using the lattice data of Ref.~\cite{cucchieri:2008}. The plot of the one- and two-loop CF predictions for the three-gluon dressing function in comparison to the lattice data of Ref.~\cite{maas:2020} is shown in \cref{fig:SU2}.

The two-loop corrections display a very similar behavior than those in the SU(3) case. More precisely, the two-loop corrections move the zero-crossing deeper in the IR. Both calculations, at one- and two-loop order reproduce very well the lattice data. The absolute error between the CF predicted three-gluon dressing function and lattice simulations is presented in \cref{tab:error-su2}.

As can be seen in \cref{tab:error-su2}, the error between the predicted three-gluon dressing function and lattice data diminishes from one- to two-loop order. As expected, the quantitative results are not as good as for the SU(3) case. This result is consistent with previous works \cite{gracey:2019,barrios:2020} in which two-loop results systematically showed a better agreement for the SU(3) gauge group. This can be attributed to the fact that the expansion parameter is larger for SU(2) in the whole range of momenta. In this case, we can also argue that because the uncertainty of lattice data is not small, a much smaller error between lattice data and the CF prediction would not make much sense.

\begin{table}[h] 
\begin{tabular}{|| c|c|c|c|c|c ||}
\hline
order & $\lambda_0$ & $m_0$ (MeV) & $\chi_{DF}(\%)$ & $\chi_{\Gamma,A}(\%)$  \\
\hline \hline
\;1-loop\; & \;\;0.42\;\; & \;\;450\;\; &\;\;7.5\;\; & \;\;15.5\;\;    \\
\hline
\;2-loop\; & \;\;0.38\;\;  & \;\;400\;\; &\;\;4.9\;\; & \;\; 12.2\;\; \\
\hline
\end{tabular}
\caption{This table shows, depending on the loop order, the values of the parameters which best fit the  lattice data of Ref.~\cite{cucchieri:2008} for the the ghost and gluon two-point functions, the corresponding joint error for the gluon and ghost dressing functions, and the individual errors for the predicted three-gluon dressing function in comparison to the data of Refs.~\cite{maas:2020}, denoted $\chi_{\Gamma,A}$.}\label{tab:error-su2}
\end{table}

\subsection{Scheme dependence}

A complementary way of testing the validity of the perturbative analysis within the CF model consists in checking the scheme dependence of the three-gluon dressing function and how it depends on the loop order. This type of analysis was performed for the two-point functions in Ref.~\cite{gracey:2019} by comparing the IRS to the vanishing momentum scheme (VM). Here, we extend this analysis to the case of the three-gluon dressing function.

The VM renormalization conditions differ from the IRS ones in that the condition $\smash{Z_{m^2}Z_A Z_c=1}$ is replaced by
\beq
G^{-1}(p=0)=\frac{1}{m^2}\,.
\eeq
Even though this scheme features a Landau pole in the IR in the case where $\mu$ is set equal to $\smash{\mu(p)=p}$ \cite{tissier:2010,tissier:2011}, we have seen that a more sensible choice is $\mu(p)=\sqrt{p^2+m_0^2}$ which stops the flow at $m_0$, and (depending on the value of $m_0$), can avoid the Landau pole. Here, we shall more generally consider the choice $\mu(p)=\sqrt{p^2+\alpha m_0^2}$ , with $\alpha=1$ or $\alpha=2$, which stops the flow at $\sqrt{\alpha}m_0$. With the goal of measuring the difference between the predicted VM and IRS three-gluon dressing functions, we introduce the following quantity
\beq
\mathcal{H}_{\alpha}=\sqrt{\frac{1}{N}\sum_{i=1}^N \left(\frac{\Gamma_{\text{VM},\alpha}(p_i)-\Gamma_{\text{IRS}}(p_i)}{\Gamma_{\text{IRS}}(p_N)}\right)^2},
\eeq
where the sum runs over the lattice points.

\cref{tab:scheme-su3,tab:scheme-su2} shows that, in all cases, the scheme dependence diminishes from one- to two-loop order, which is consistent with the perturbative paradigm within the CF model. Moreover, we observe that the scheme dependence is stronger in the SU(2) case, which is in line with previous observations. 

\begin{table}[t] 
\begin{tabular}{|| c|c|c|c|c ||}
\hline
order & $\mathcal{H}_{\alpha=1,\text{A}}(\%)$ & $\mathcal{H}_{\alpha=2,\text{A}}(\%)$ & $\mathcal{H}_{\alpha=1,\text{C}}(\%)$ & $\mathcal{H}_{\alpha=2,\text{C}}(\%)$   \\
\hline \hline
\;1-loop\; & \;\;5.1 \;\; & \;\; 5.3 \;\; & \;\; 2.4 \;\; & \;\; 2.5 \;\;  \\
\hline
\;2-loop\; & \;\; 3.1 \;\;  & \;\; 2.9 \;\; & \;\; 1.6 \;\; & \;\; 1.5 \;\; \\
\hline
\end{tabular}
\caption{\label{tab:scheme-su3} Scheme dependence in the SU(3) case. The normalization of the three-gluon dressing function was chosen so as to minimize the disagreement with lattice simulations from Ref.~\cite{aguilar:2021}, in the case of $\mathcal{H}_{\alpha,\text{A}}$, and with lattice data from Ref.~\cite{catumba:2022}, in the case of $\mathcal{H}_{\alpha,\text{C}}$.}
\end{table}

\begin{table}[t] 
\begin{tabular}{|| c|c|c ||}
\hline
order & $\mathcal{H}_{\alpha=1,\text{A}}(\%)$ & $\mathcal{H}_{\alpha=2,\text{A}}(\%)$    \\
\hline \hline
\;1-loop\; & \;\;11.6 \;\; & \;\; 11.3 \;\;  \\
\hline
\;2-loop\; & \;\; 5.1 \;\;  & \;\; 5.7 \;\; \\
\hline
\end{tabular}
\caption{\label{tab:scheme-su2} Scheme dependence in the SU(2) case. The normalization of the three-gluon dressing function was chosen so as to minimize the disagreement with lattice simulations from Ref.~\cite{maas:2020}.}
\end{table}

\section{Conclusions}
Using the Curci-Ferrari model, we have evaluated one of the dressing functions that occur in the three-gluon vertex function as one of the external momenta is taken to zero. Our two-loop calculation extends earlier one-loop results \cite{pelaez:2013} in that particular configuration of momenta. Since the parameters of the model were fixed from a similar analysis of the two-loop two-point functions \cite{gracey:2019}, our results appear as a pure prediction of the model which can be compared to lattice data both in the SU(2) case and in the SU(3) case. We find that the two-loop corrections systematically improve the comparison to the data. This is particularly true in the SU(3) case which we interpret as originating in the fact that the coupling constant is smaller in this case than in the SU(2) case. We also find that scheme dependences get reduced at two-loop order, specially in the SU(3) case. All these results reinforce the idea that certain quantities in YM theory are akin to perturbative methods, through the CF model.

We have also provided a detailed analysis of the IR structure of the CF model which allows one to unveil the dominant contributions to each vertex function and their diagrammatic origin. Thanks to this analysis, we could derive an exact formula for the leading order IR asymptotic behavior of the three-gluon dressing in the form of the one-loop result multiplied by the cube of the ghost dressing function at zero momentum. Similar formulas apply to the two- and four-point functions. In the case, of the three-gluon dressing, it shows that zero-crossing does occur in the CF model. However, our fit to the data shows that the scale of the zero-crossing is considerably reduced when going from one- to two-loop order. This result seems compatible with certain recent lattice simulations.

\section*{Acknowledgments}
We are grateful to A. Maas, O. Oliveira and J. Rodr\'iguez-Quintero for kindly sharing the data of Refs. \cite{maas:2020}, \cite{catumba:2022} and \cite{aguilar:2021}. We also would like to thank J.~A. Gracey, J.~Serreau, M.~Tissier and N.~Wschebor for collaboration on related works and for useful insight on the manuscript. Also, our discussion of the IR structure of the Curci-Ferrari model largely benefited from interactions with M.~Tissier. N.~B. acknowledges the financial support from the PEDECIBA program, the ``Comisi\'on Acad\'emica de Posgrado" (CAP) as well as the ``Institut Franco-Uruguayen de Physique'' (Laboratoire International Associ\'e, CNRS). 

\appendix

\section{Diagrams}\label{ap:diagrams}
\begin{figure}[t]
\centering
\begin{subfigure}[b]{.11\textwidth}
\centering
\includegraphics[width=\textwidth]{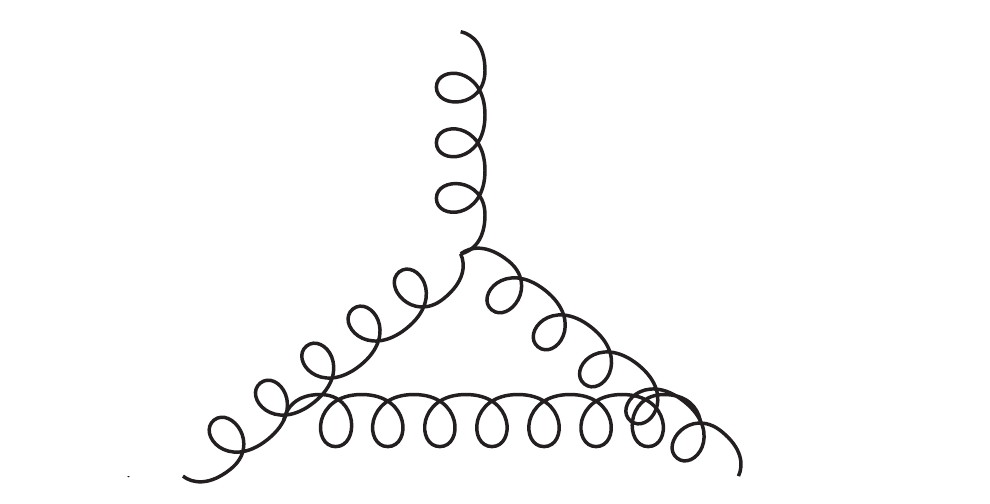} 
\end{subfigure} 
\begin{subfigure}[b]{.11\textwidth}
\centering
\includegraphics[width=\textwidth]{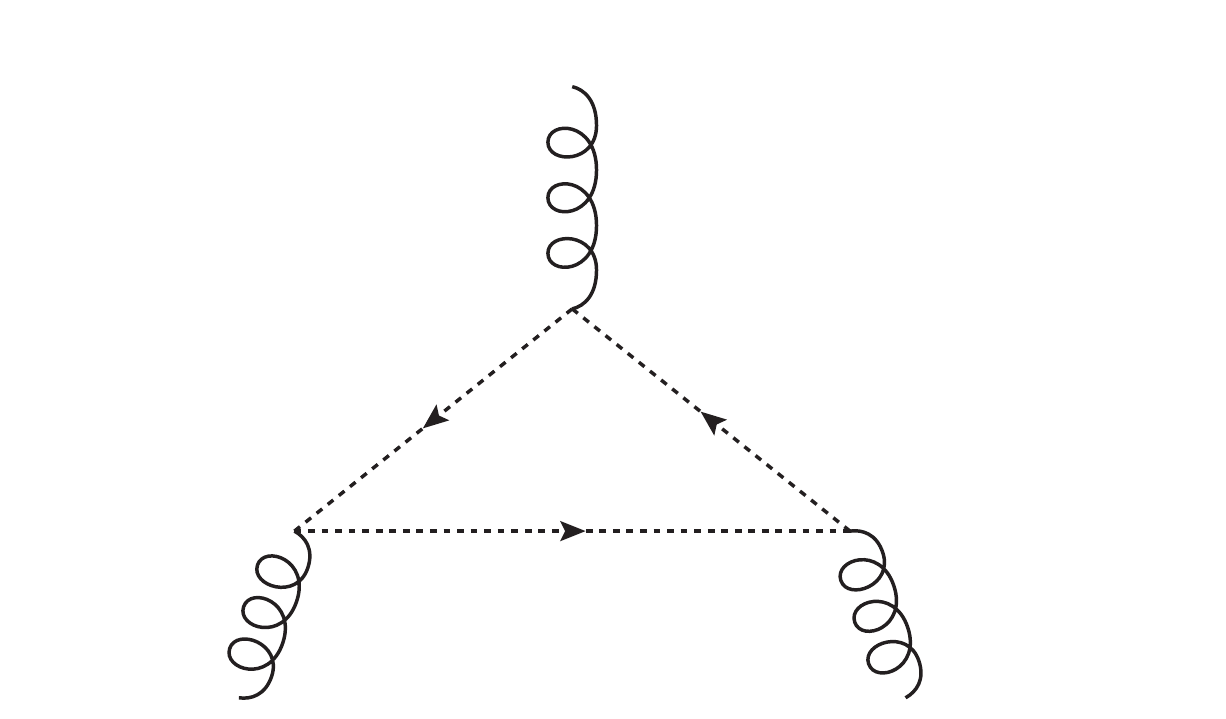} 
\end{subfigure} 
\begin{subfigure}[b]{.11\textwidth}
\centering
\includegraphics[width=\textwidth]{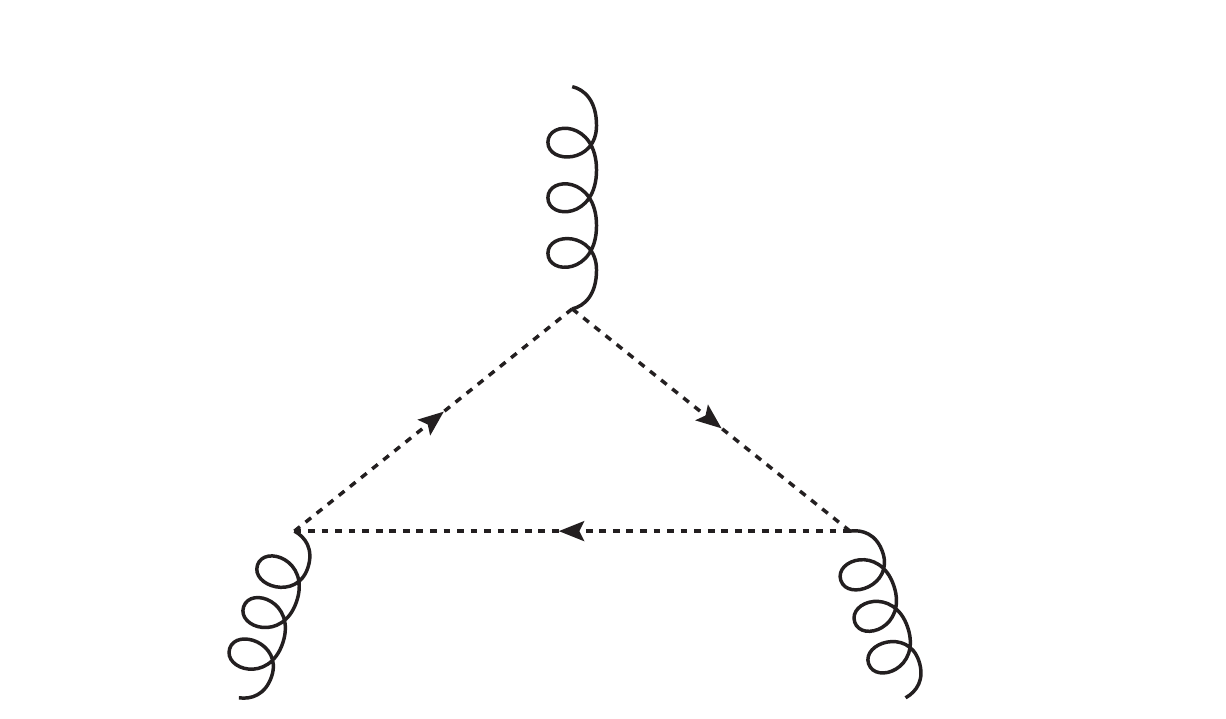} 
\end{subfigure} 
\begin{subfigure}[b]{.10\textwidth}
\includegraphics[width=\textwidth]{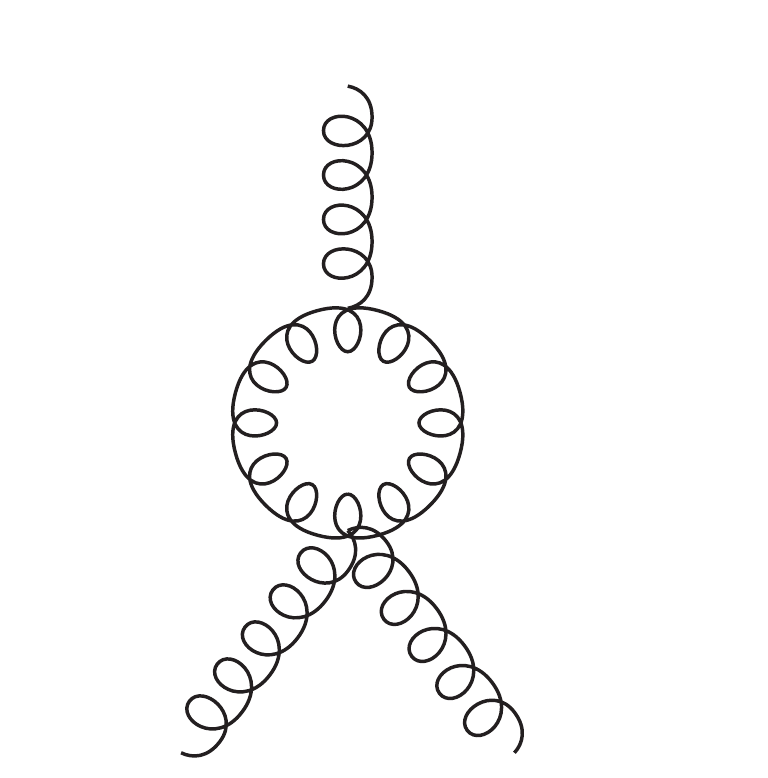} 
\end{subfigure} 
\caption{One-loop diagrams contributing to $\Gamma(p^2)$.}\label{fig:1loop_diag}
\end{figure}

\begin{figure}[t]
\centering
\begin{subfigure}[b]{.12\textwidth}
\centering
\includegraphics[width=\textwidth]{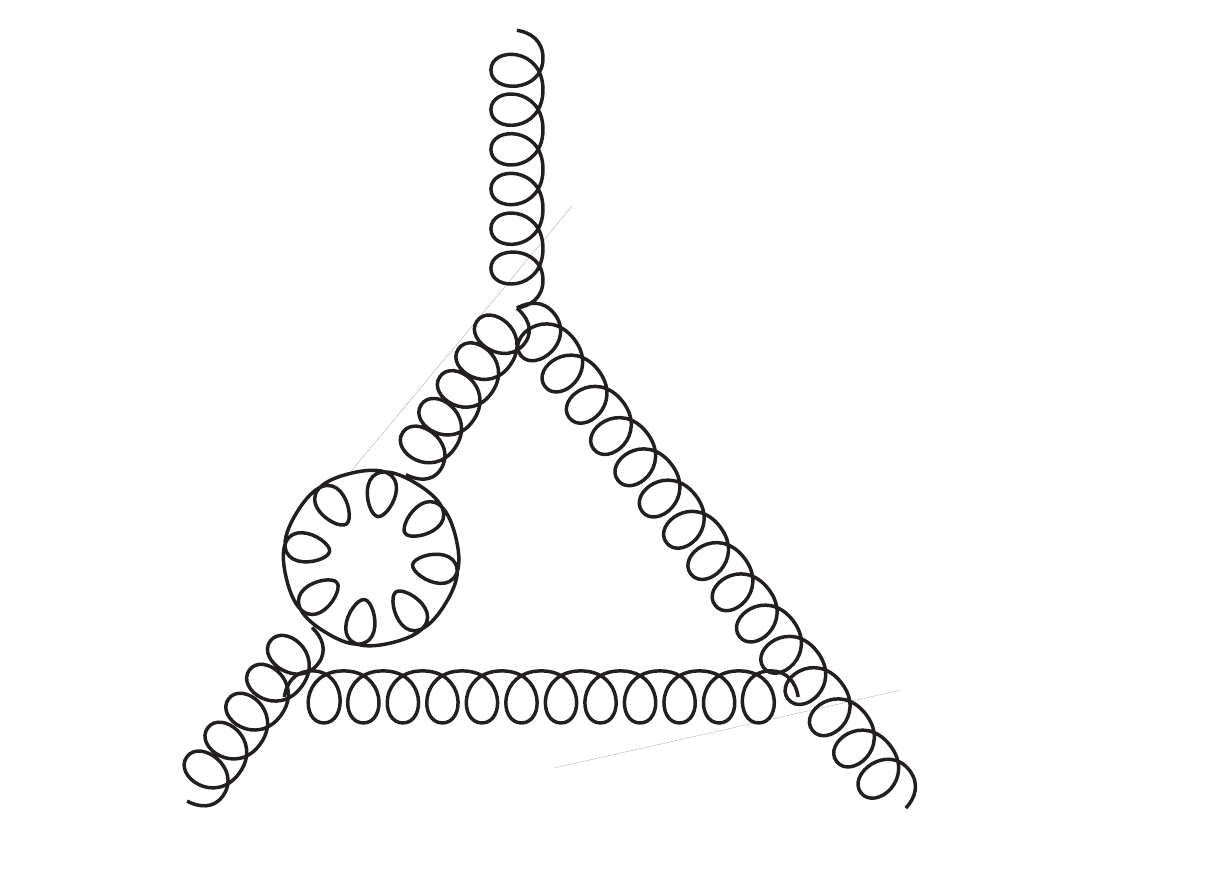} 
\end{subfigure}
\begin{subfigure}[b]{.12\textwidth}
\centering
\includegraphics[width=\textwidth]{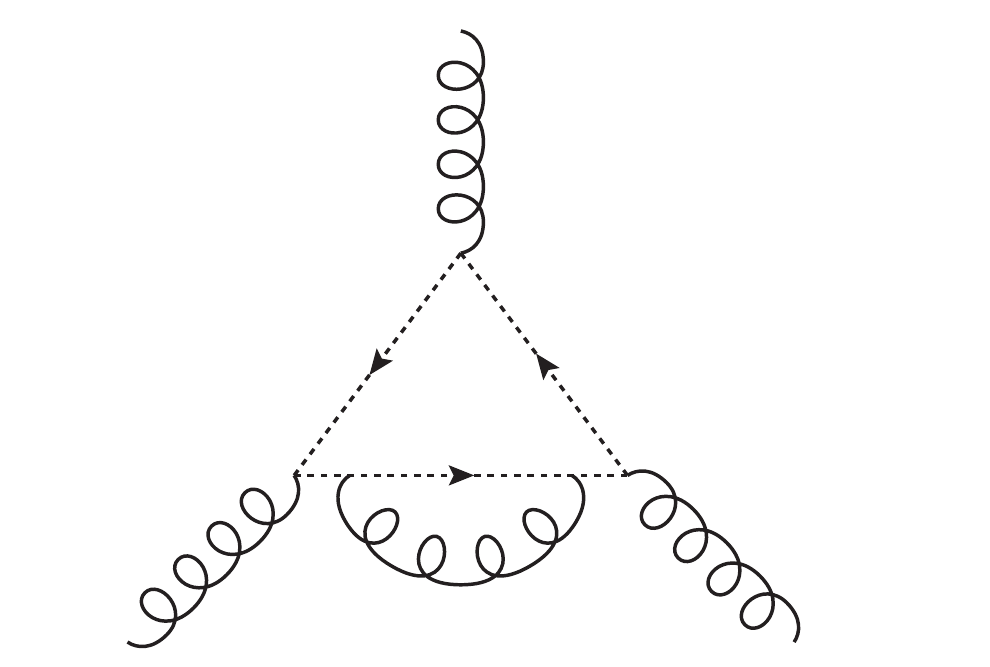} 
\end{subfigure}
\begin{subfigure}[b]{.12\textwidth}
\centering
\includegraphics[width=\textwidth]{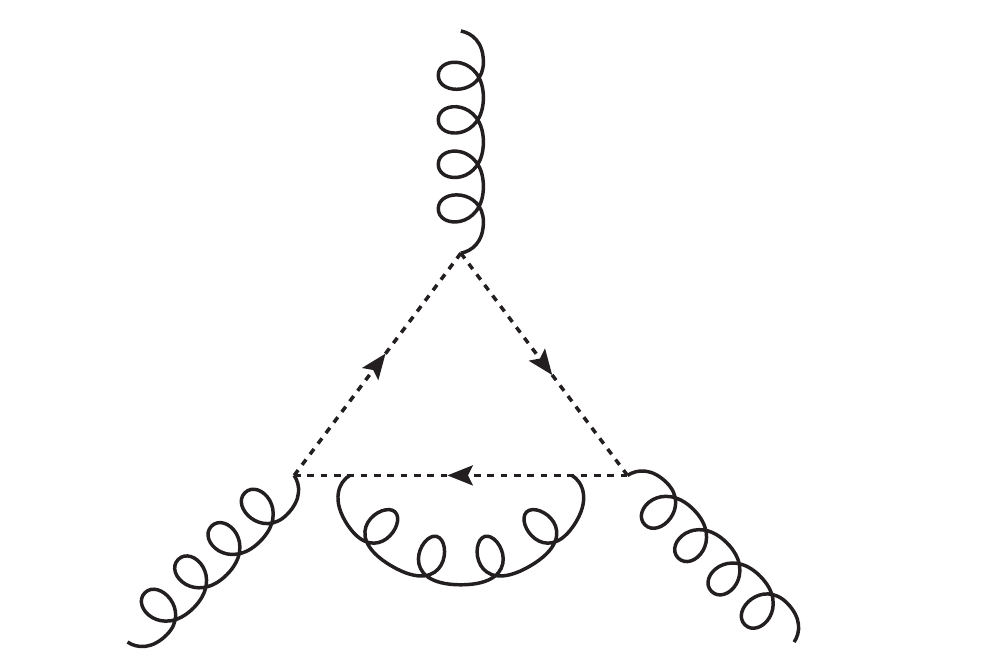} 
\end{subfigure}
\begin{subfigure}[b]{.12\textwidth}
\centering
\includegraphics[width=\textwidth]{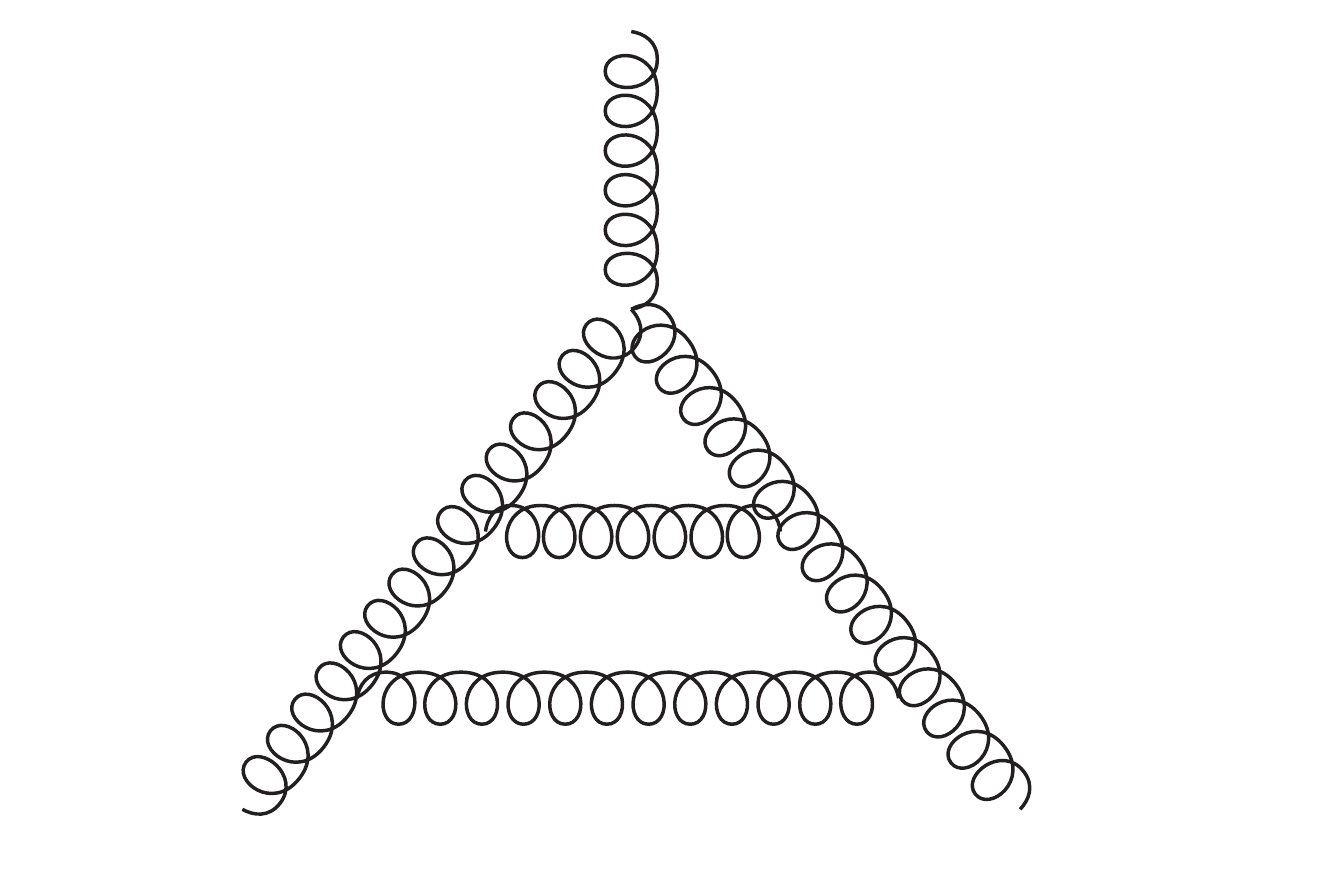} 
\end{subfigure}
\begin{subfigure}[b]{.12\textwidth}
\centering
\includegraphics[width=\textwidth]{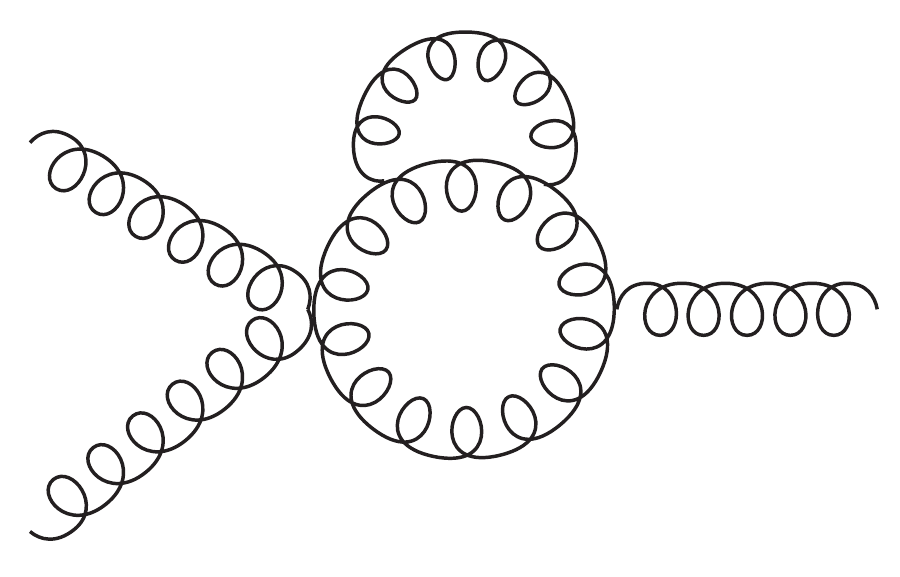} 
\end{subfigure}
\begin{subfigure}[b]{.12\textwidth}
\centering
\includegraphics[width=\textwidth]{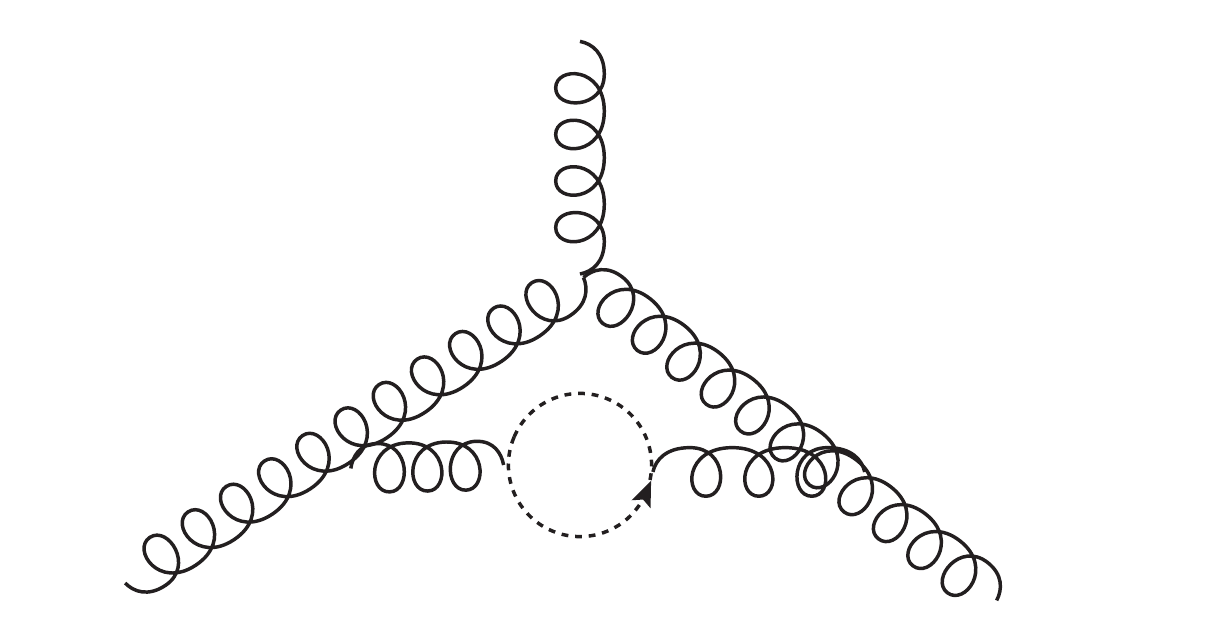} 
\end{subfigure}
\begin{subfigure}[b]{.12\textwidth}
\centering
\includegraphics[width=\textwidth]{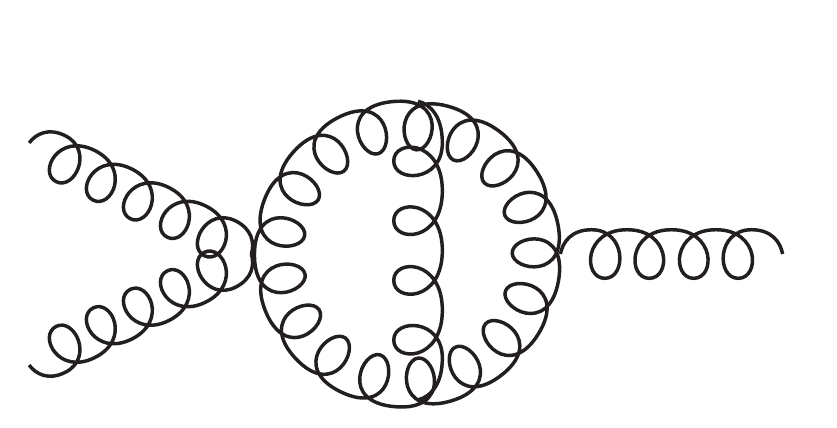} 
\end{subfigure}
\begin{subfigure}[b]{.12\textwidth}
\centering
\includegraphics[width=\textwidth]{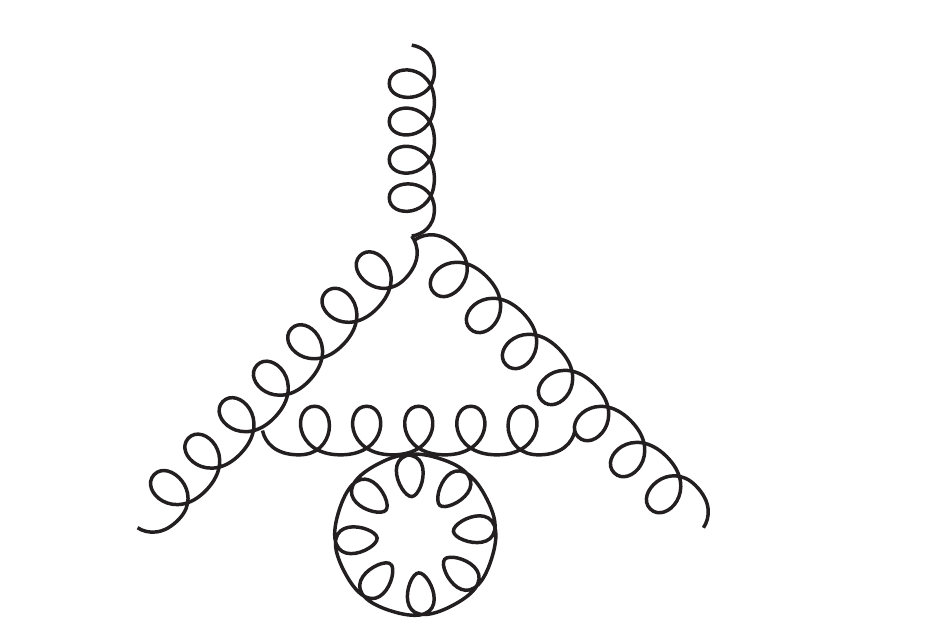} 
\end{subfigure}
\begin{subfigure}[b]{.12\textwidth}
\centering
\includegraphics[width=\textwidth]{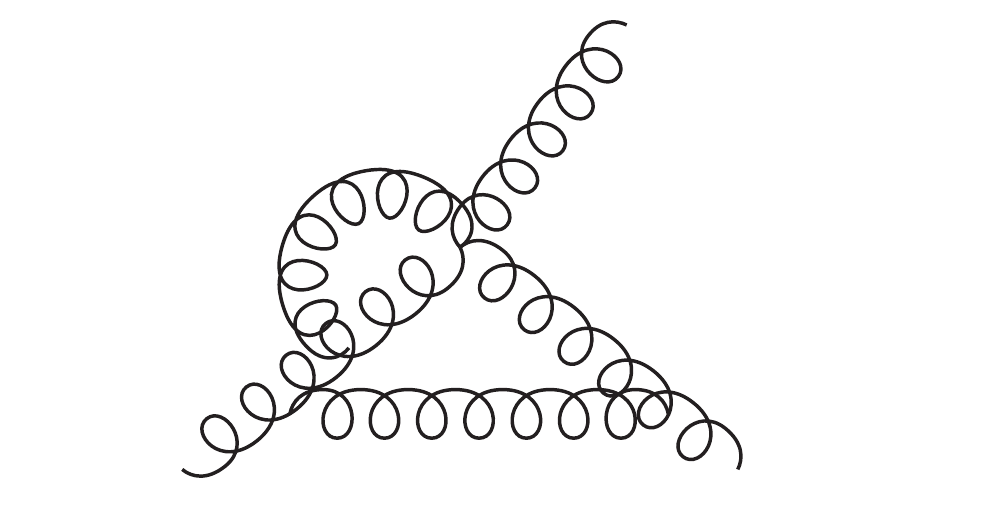} 
\end{subfigure}
\begin{subfigure}[b]{.12\textwidth}
\centering
\includegraphics[width=\textwidth]{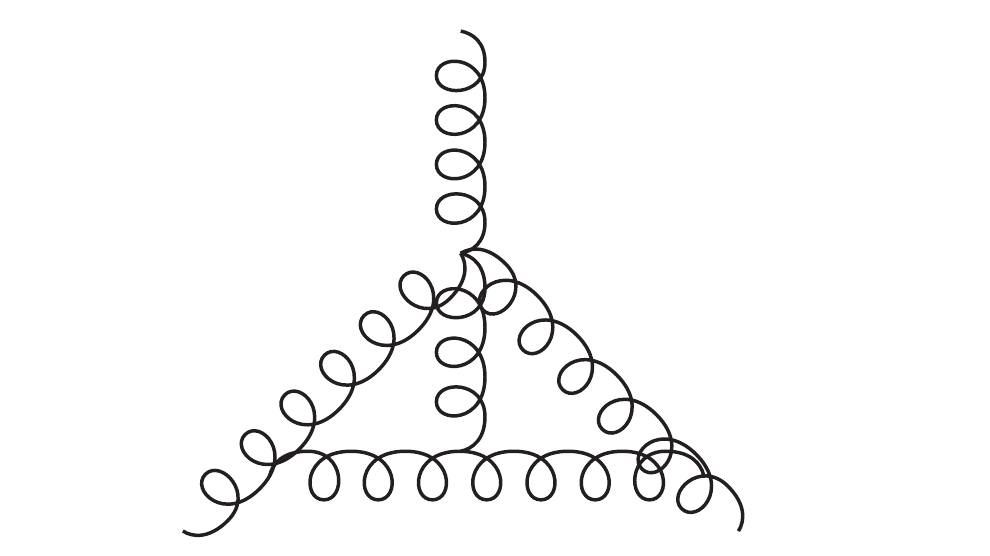} 
\end{subfigure}
\begin{subfigure}[b]{.12\textwidth}
\centering
\includegraphics[width=\textwidth]{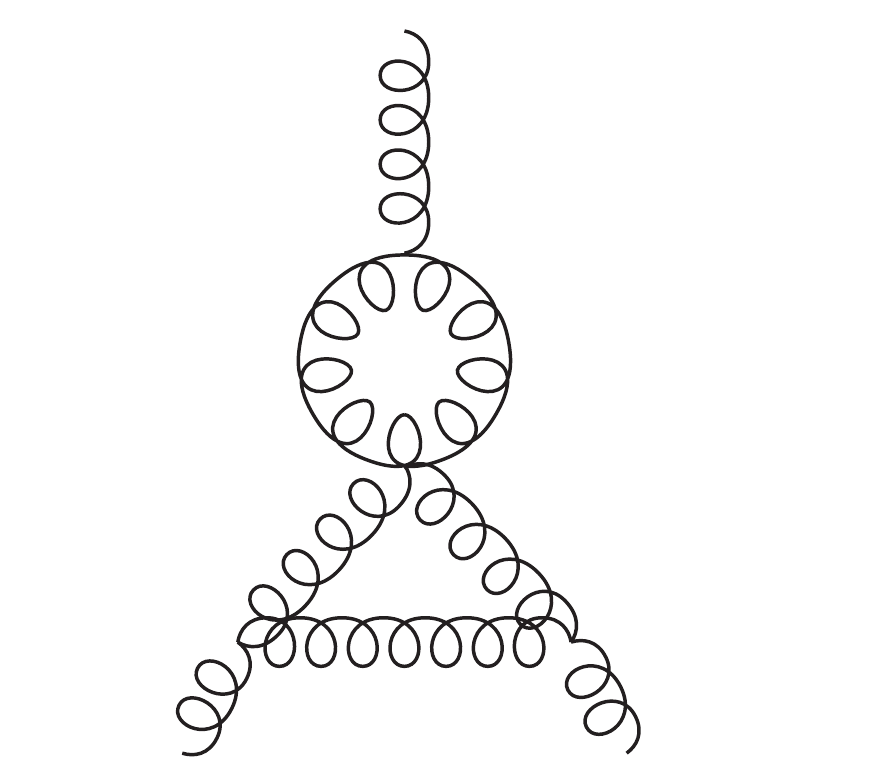} 
\end{subfigure}
\begin{subfigure}[b]{.12\textwidth}
\centering
\includegraphics[width=.5\textwidth]{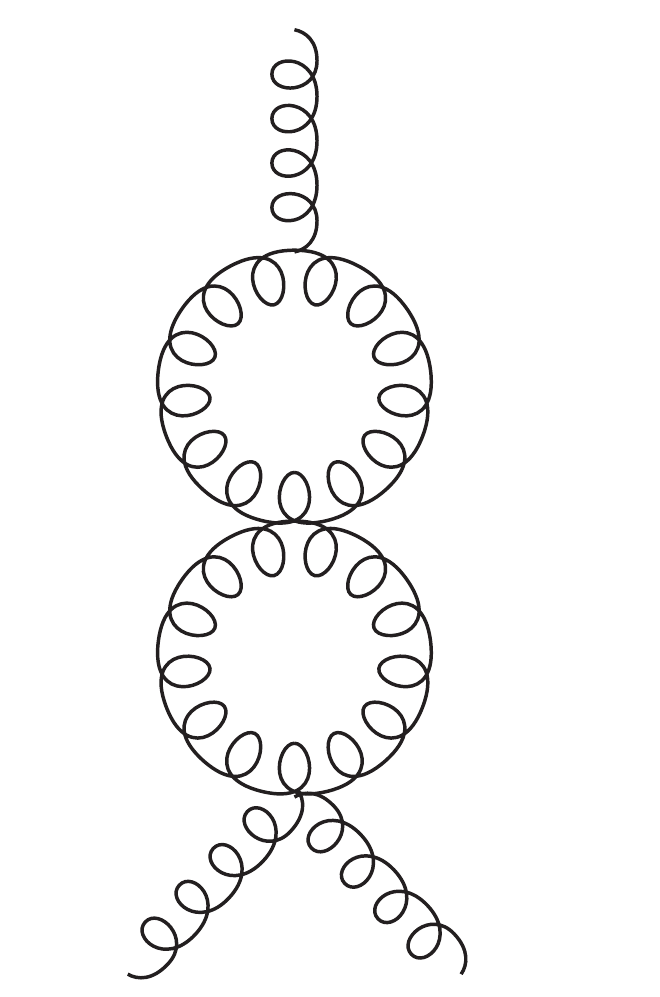} 
\end{subfigure}
\begin{subfigure}[b]{.12\textwidth}
\centering
\includegraphics[width=\textwidth]{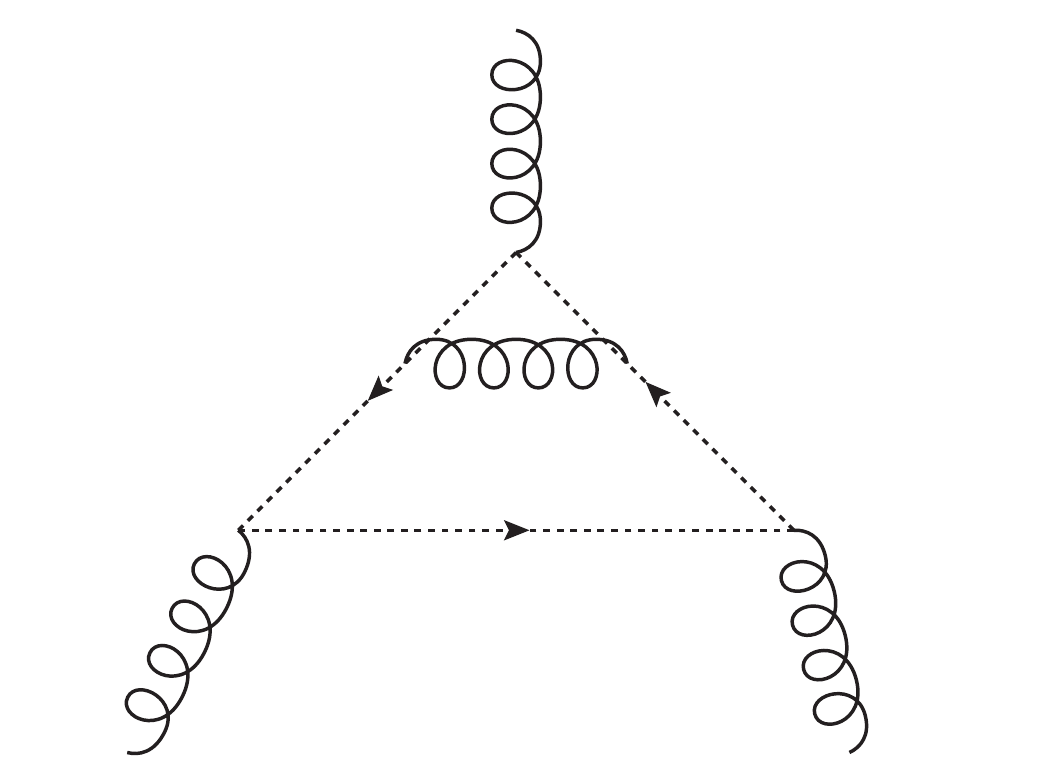} 
\end{subfigure}
\begin{subfigure}[b]{.12\textwidth}
\centering
\includegraphics[width=\textwidth]{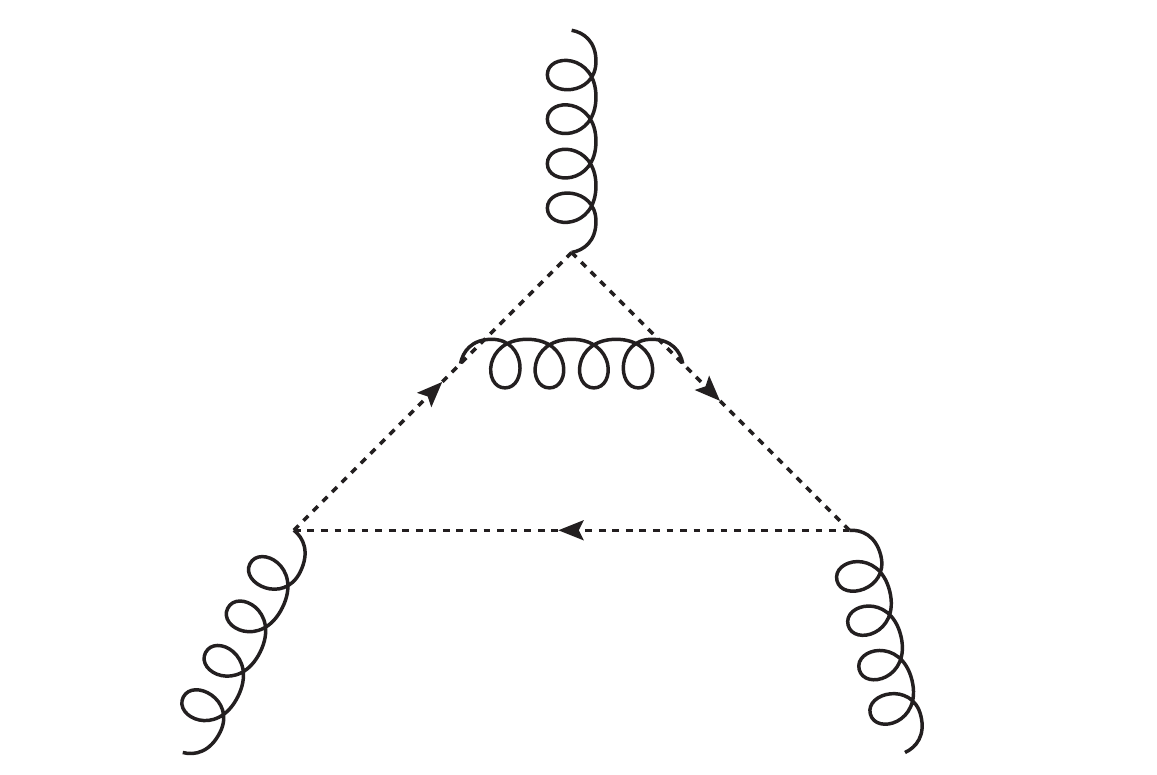} 
\end{subfigure}
\begin{subfigure}[b]{.12\textwidth}
\centering
\includegraphics[width=.7\textwidth]{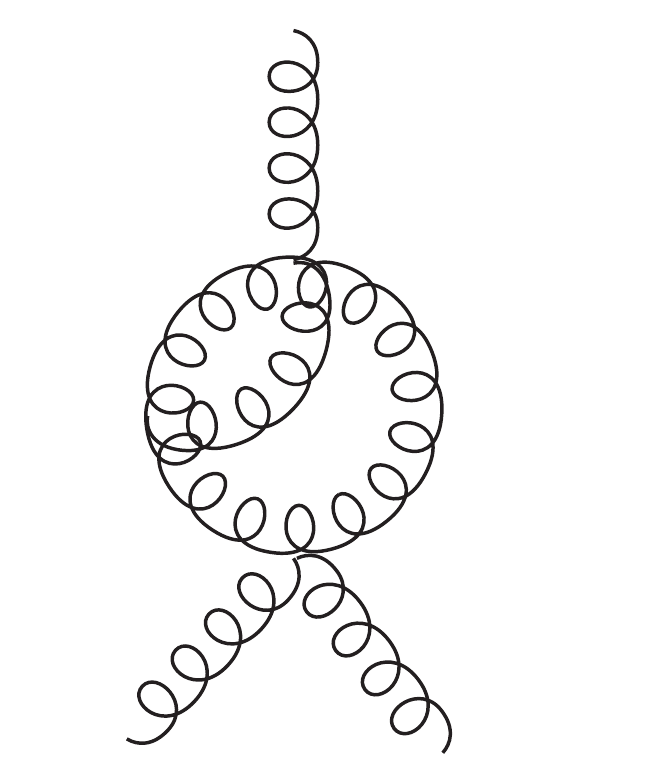} 
\end{subfigure}
\begin{subfigure}[b]{.12\textwidth}
\centering
\includegraphics[width=\textwidth]{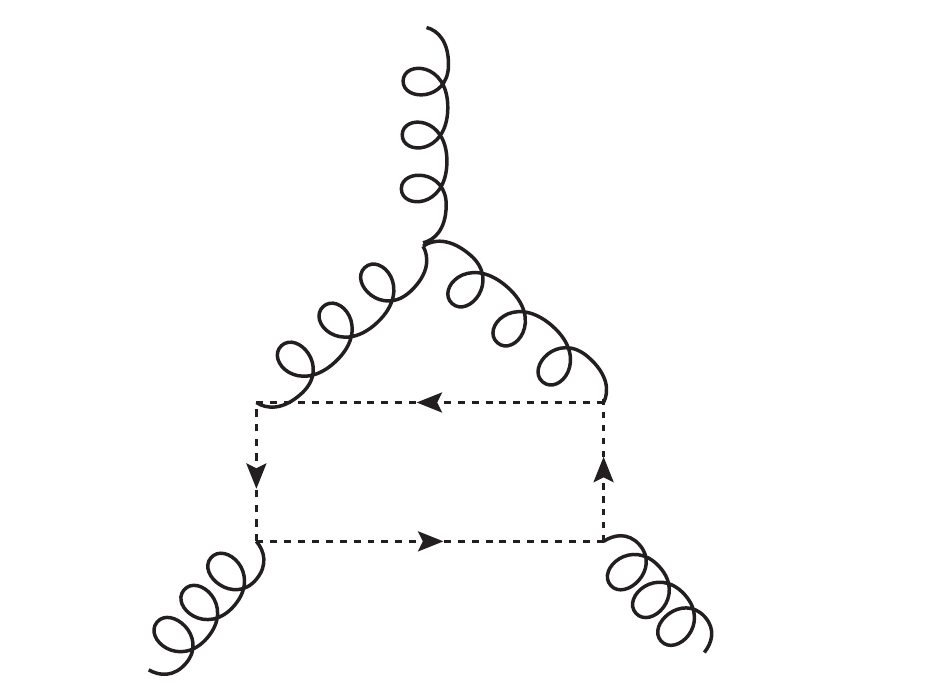} 
\end{subfigure}
\begin{subfigure}[b]{.12\textwidth}
\centering
\includegraphics[width=\textwidth]{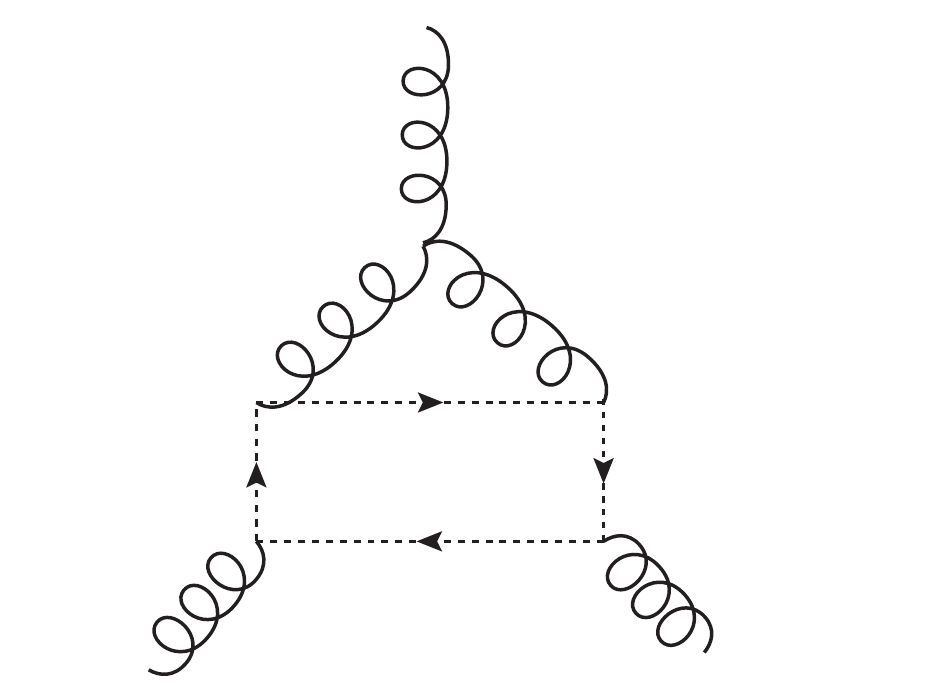} 
\end{subfigure}
\begin{subfigure}[b]{.12\textwidth}
\centering
\includegraphics[width=\textwidth]{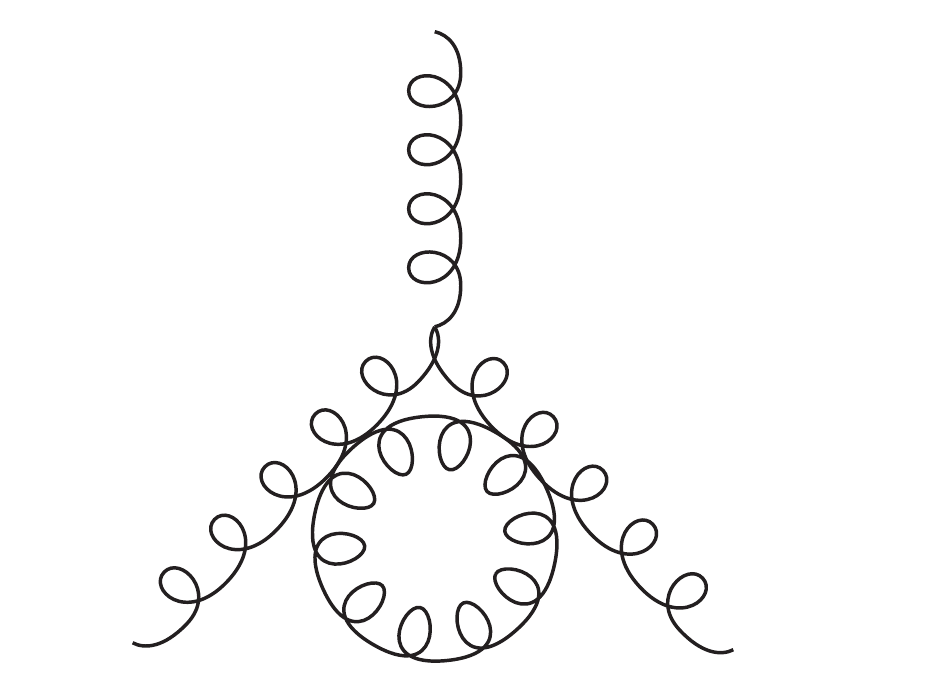} 
\end{subfigure}
\begin{subfigure}[b]{.12\textwidth}
\centering
\includegraphics[width=1.1\textwidth]{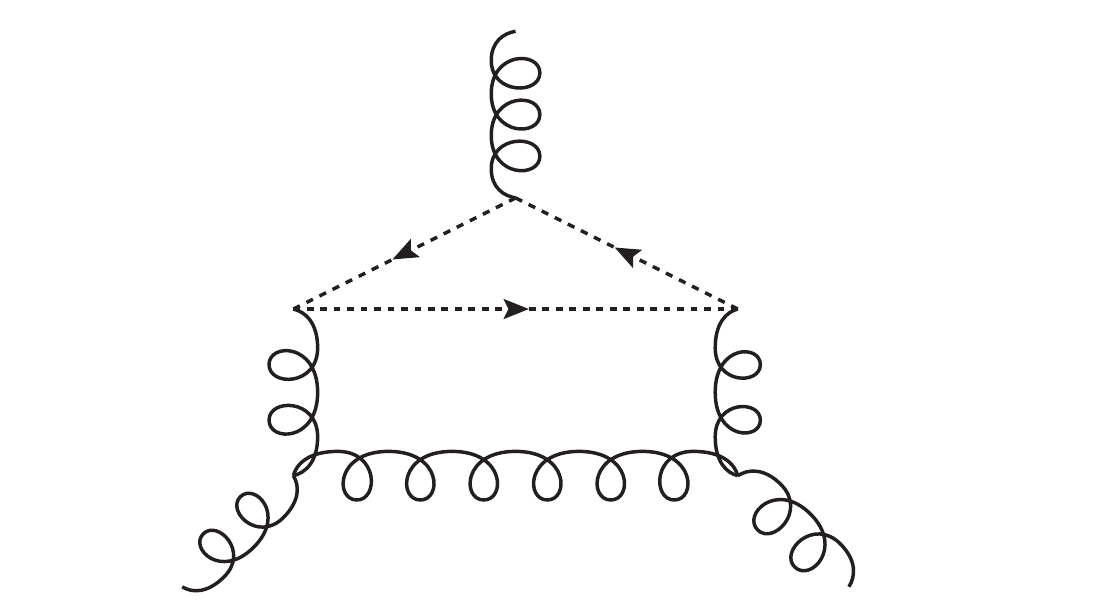} 
\end{subfigure}
\begin{subfigure}[b]{.12\textwidth}
\centering
\includegraphics[width=1.1\textwidth]{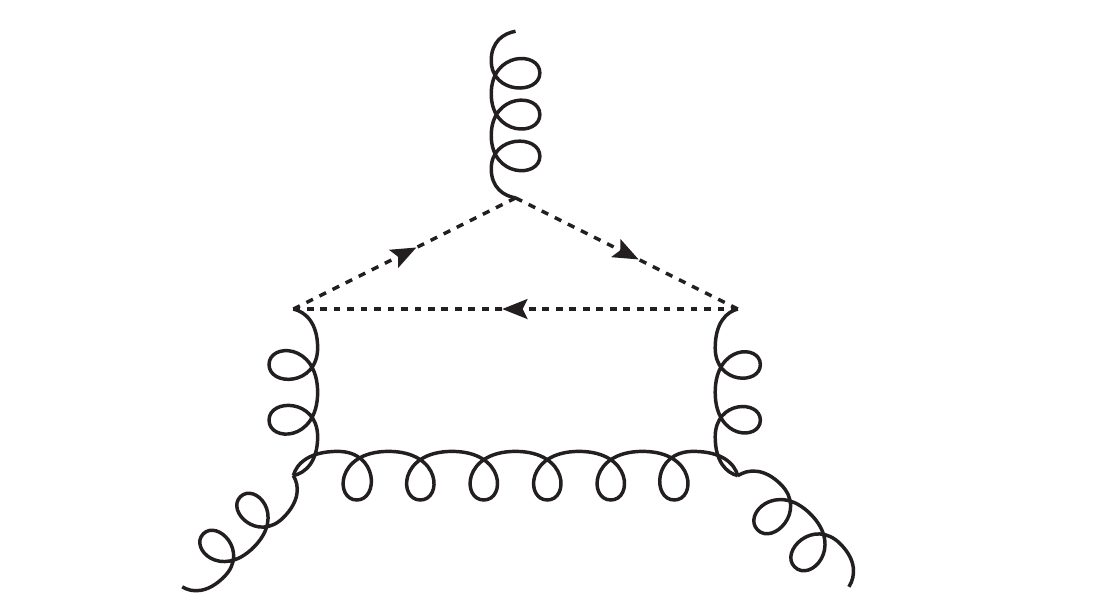} 
\end{subfigure}
\begin{subfigure}[b]{.12\textwidth}
\centering
\includegraphics[width=.9\textwidth]{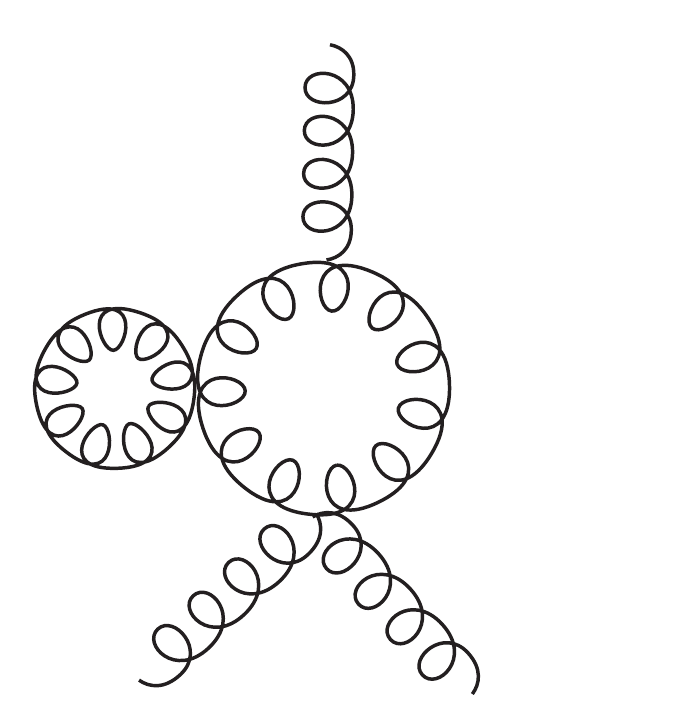} 
\end{subfigure}
\begin{subfigure}[b]{.12\textwidth}
\centering
\includegraphics[width=.8\textwidth]{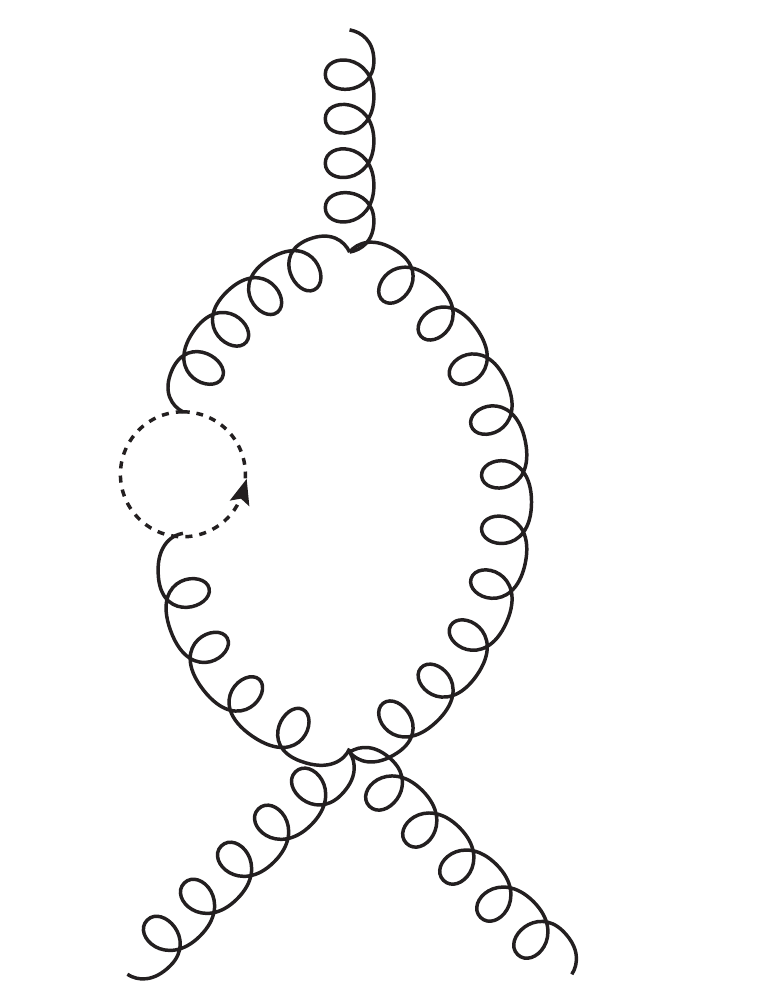} 
\end{subfigure}
\begin{subfigure}[b]{.12\textwidth}
\centering
\includegraphics[width=.9\textwidth]{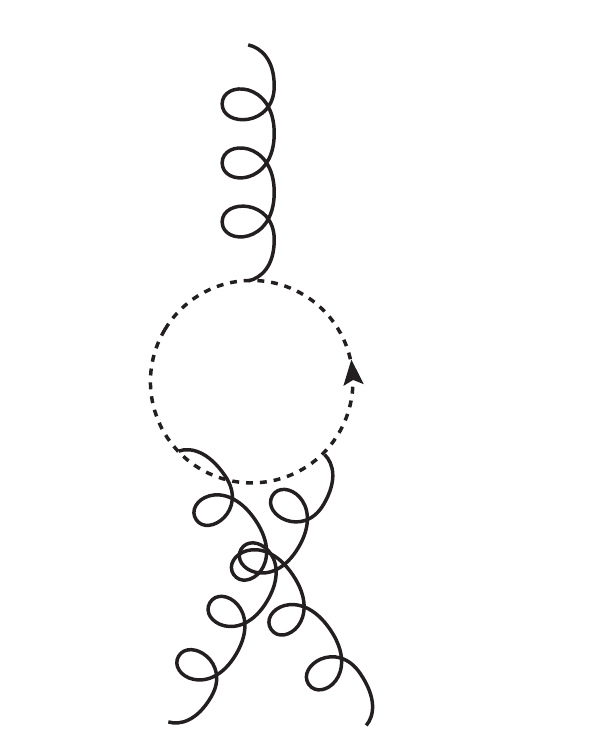} 
\end{subfigure}
\caption{Two-loop diagrams contributing to $\Gamma(p^2)$.}\label{fig:2loop_diag}
\end{figure}

 All Feynman diagrams in this section and in the rest of the article have been drawn with \textsc{Jaxodraw} \cite{binosi:2009}. The one-loop diagrams contributing to $\Gamma(p^2)$ are shown in Fig.~\ref{fig:1loop_diag} while the two-loop diagrams are shown in Fig.~\ref{fig:2loop_diag}. We do not draw diagrams which are permutations of the ones below. Also, as already noticed in the main text, we do not draw the non-planar diagrams which are zero due to a vanishing color factor.
 
\section{IR expansion of Feynman graphs}\label{app:AI}

In the next section, we provide a detailed study of the IR structure of the CF model. The analysis is based on the large mass expansion of Feynman graphs in terms of asymptotically irreducible subgraphs \cite{Smirnov:2002pj}, as well as on the notions of Taylor and asymptotic mass powers. In the present section, preparing the ground for this analysis, we introduce these notions in the case of a scalar theory involving one massive field of mass $m$ and one massless field.

\subsection{Asymptotically irreducible subgraphs}
Consider a graph ${\cal G}$ and denote $G(p_i,m)$ the associated Feynman integral. We are interested in the regime where all the external momenta $p_i$ are much smaller than $m$. One could first start by considering a naive infrared expansion of $G(p_i,m)$ obtained by Taylor expanding the corresponding integrand in powers of the external momenta $p_i$. Except for very specific cases, however, this Taylor expansion, denoted ${\cal T}_{p_i}G(p_i,m)$, is not the actual asymptotic expansion of the graph, denoted ${\cal A}_{p_i}G(p_i,m)$. The reason is that, in the case where there is no way of routing the external momenta such that they avoid the massless lines of ${\cal G}$, the Taylor expansion of these external momenta eventually produces infrared divergences. Although the latter are regularized within dimensional regularization, their appearance is symptomatic of the invalidity of the Taylor expansion beyond a certain order.

The correct asymptotic expansion is obtained from the following formula \cite{Smirnov:2002pj}
\beq
{\cal A}_{p_i} G(p_i,m)=\!\!\!\!\!\!\sum_{\bar {\cal G}(AI)\subset\,{\cal G}} \int_{q_j} R(q_j,p_i) \, {\cal T}_{q_j,p_i}\bar G(q_j,p_i,m)\,,\label{eq:sum}
\eeq
where the sum is to be taken over the {\it asymptotically irreducible (AI) subgraphs} $\bar {\cal G}$ of the original graph with associated Feynman integral $\bar G(q_j,p_i,m)$.  An asymptotically irreducible subgraph is defined as any subgraph that contains all the massive lines of the original graph and that is one-particle-irreducible (1PI) with respect to the massless lines. Basically, it represents one possible way to route large momenta through some of the massless lines of the original graph. This allows one to safely Taylor-expand $\bar G(q_j,p_i,m)$ with respect to all the external momenta of the subgraph.\footnote{We should stress here that the Taylor expansion of a given AI subgraph is not, in general, its actual asymptotic expansion. It should be considered more as a bookkeeping device that helps organizing the asymptotic expansion of the original graph and that makes sense only after the contributions from all the AI subgraphs have been added. Similarly, the contribution associated to an AI subgraph needs to be considered as one in which the momenta flowing through the massless lines of the AI subgraph are considered large. Although this would require, in principle, to split the loop momenta into large and small, it can be argued that this is not necessary in dimensional regularization and that a formula such as (\ref{eq:sum}) does not over-count contributions.} The reason why all the massive lines are included within the AI subgraphs is that they can always be safely Taylor-expanded. It is also important to note that, in the case where the diagram involves only massless propagators, there is only one term in the asymptotic expansion (\ref{eq:sum}) corresponding to the diagram itself. One could interprete this by saying that there is only one AI subgraph to be considered in this case, the empty subgraph.\footnote{In fact, other AI subgraphs would contribute to $0$ in this case since their Taylor expansion vanishes in dimensional regularization. Also, in the case of diagrams containing some massive lines, the empty AI subgraph does not enter (\ref{eq:sum}) since the massive lines need always to be included within any AI subgraph.}

We stress that a given AI subgraph is not necessarily connected. However, its connected components appear as trees of (connected) 1PI subgraphs with respect to all types of lines, linked to each other by massive lines. In what follows, we refer to the 1PI subgraphs and the massive lines as the nodes and the branches of the tree (and, by extension, of the AI subgraph), respectively.

\subsection{Taylor and asymptotic mass powers}

The formula (\ref{eq:sum}) needs to be seen as a large mass expansion, valid insofar all external momenta $p_i$ are small with respect to $m$, irrespectively of how these momenta are related to each other. In order to organize tha various terms of the large mass expansion while unveiling their diagrammatic origin, it is convenient to introduce the notion of Taylor and asymptotic mass powers.

First, the Taylor expansion of an AI subgraph $\bar {\cal G}$ leads to a sequence of terms of the form $m^{\omega}P^n$, with $n$ a positive integer and where $P^n$ is a shorthand notation for a degree $n$ monomial in the components of the external momenta of the subgraph (which include the $p_i$ and $q_j$ above). The exponent $\omega$ will be referred to as the {\it Taylor mass power} of the corresponding term of the expansion. It is such that $\smash{\omega+n=\delta_{\bar {\cal G}}}$ where $\delta_{\bar {\cal G}}$ denotes the mass dimension of the subgraph. Since $\smash{n\geq 0}$, we have $\omega\leq\delta_{\bar {\cal G}}$. Moreover, increasing orders of the Taylor expansion appear according to decreasing Taylor mass powers. The leading term of the expansion corresponds then to the term with highest Taylor mass power $\omega_{\bar {\cal G}}$, referred to as the {\it Taylor mass power of the subgraph.} If no symmetries are present, we will typically find $\smash{\omega_{\bar {\cal G}}=\delta_{\bar {\cal G}}}$. In contrast, symmetries may decrease the value of $\omega_{\bar {\cal G}}$ strictly below $\delta_{\bar {\cal G}}$, as we will see explicitly in the next section. 

Second, from the above considerations as well as from Eq.~(\ref{eq:sum}), it is pretty clear that the asymptotic expansion of the original graph also leads to a sequence of terms of the form $m^\nu H_\alpha(p_i)$, with $H_\alpha(p_i)$ an homogeneous function of the external momenta of degree $\alpha$,  with $\smash{\nu+\alpha=\delta_{{\cal G}}}$. As opposed to $n$ above, $\alpha$ is not restricted to positive values and so $\nu$ can be strictly larger than $\delta_{{\cal G}}$. The exponent $\nu$ will be referred to as the {\it asymptotic mass power} of the corresponding term in the expansion. As before, increasing orders of the asymptotic expansion appear according to decreasing asymptotic mass powers. The leading term of the expansion corresponds to the term with highest asymptotic mass power $\nu_{{\cal G}}$, referred to as the {\it asymptotic mass power of the graph.}\footnote{Taylor and asymptotic mass powers (as well as $\delta_{\bar {\cal G}}$, $\delta_{{\cal G}}$ or $\alpha$) are integers modulo terms proportional to $\epsilon$. In most of the subsequent discussion, the $\epsilon$-dependent part will not play any role and we shall treat mass powers as integers, neglecting their $\epsilon$-dependent part. The latter will play a role when investigating the origin of the logarithms in Sec.~\ref{sec:logs}.}

The notions of asymptotic and Taylor mass powers are very useful for unveiling the origin of the various terms in the large mass asymptotic expansion of a given graph ${\cal G}$. Indeed, the Taylor mass power $\omega_{\bar {\cal G}}$ of an AI subgraph $\bar {\cal G}$ gives the highest asymptotic mass power it contributes to in the asymptotic expansion of the original graph. In particular, the leading terms in the asymptotic expansion are those corresponding to AI subgraphs with highest Taylor mass power, Taylor-expanded to leading order. The next-to-leading terms come either from the AI subgraphs with highest Taylor mass power, Taylor-expanded to next-to-leading order, or from the AI subgraphs with next-to-highest Taylor mass power Taylor-expanded to leading order, and so on and so forth.

\section{IR structure of the CF model}\label{app:AI_CF}
We now extend and apply the previous considerations to the case of the CF model.

\subsection{AI subgraphs in the CF model}

The CF model features massless ghosts and massive gluons. We note, however, that the CF gluon propagator writes as
\beq
\frac{P^\perp_{\mu\nu}(q)}{q^2+m^2}=\frac{1}{q^2+m^2}\left[\delta_{\mu\nu}+\frac{q_\mu q_\nu}{m^2}\right]-\frac{1}{m^2}\frac{q_\mu q_\nu}{q^2}\,,\label{eq:prop}
\eeq 
so it contains both a massive and a ``massless'' component. A priori, one should take this fact into account when identifying the AI subgraphs. To do so, one should first imagine replicating the graph by deciding which gluon lines carry massive components of the gluon propagator and which ones carry massless components. For any such choice, the AI subgraphs should contain all gluon lines except some of those corresponding to massless components, and form trees whose branches are some of the gluon lines corresponding to massive components. Now, for a given choice of gluon lines that are left out of the AI subgraph (because they correspond to massless components) and a given choice of gluon lines that form the branches of the tree (because they correspond to massive components), one finds all possible choices of gluon components inside the nodes of the tree and these gluon components reconstruct the fully transverse propagator (\ref{eq:prop}). 

Beyond its relevance in the argumentation below, this remark leads to a simple procedure to obtain all AI subgraphs of a given graph in the CF model. One first leaves out certain gluon lines. Then one looks for all the graphs that contain the other gluon lines and whose connected components form trees whose branches are some of these gluon lines. The gluon lines inside the nodes of these trees correspond to the fully transverse propagator (\ref{eq:prop}), while the branches of the tree correspond to massive components, and finally, the gluon lines that were left out correspond to massless components. We note that the latter contribute with a factor $1/m^2$ each. Even though this factor should be considered as part of $R(q_j,p_i)$ in Eq.~(\ref{eq:sum}), we shall conveniently absorb it within the other factor, so that $R(q_j,p_i)$ remains independent of the mass. We stress once more that, in the case where all gluon lines have been interpreted as the massless component, the diagram contains only massless propagator (including possible ghost propagators) and the only relevant AI subgraph is the empty one, meaning that the integral should be considered unexpanded in Eq.~(\ref{eq:sum}).

\subsection{Taylor mass power of an AI subgraph}\label{sec:Taylor}
We now would like to evaluate the Taylor mass power of an AI subgraph occurring in the CF model. To this purpose, we denote by $\bar {\cal G}_i$ the various nodes of $\bar {\cal G}$ and by $I_A$ the total number of gluon lines letting aside those that are hidden in the nodes. Because each of these gluon lines contributes a term $-2$ to the Taylor mass power of the AI subgraph, we find
\beq
\omega_{\bar {\cal G}}=-2I_A+\sum_i \omega_{\bar {\cal G}_i}\,.\label{eq:w}
\eeq
Recall that we have decided to define $R(q_j,p_i)$ such that it does not depend on $m$, so even though some of the gluon lines corresponding to massless components are not included in the AI subgraphs, the associated factors $1/m^2$ contribute to the evaluation of the Taylor mass power. In particular, the formula includes the case where all the gluon lines are attributed the massless component. In this case, the AI subgraph is empty but because each gluon lines contributes a trivial factor $1/m^2$, we have $\omega_{\bar {\cal G}}=-2I_A$.

From Eq.~(\ref{eq:w}), we are led to the determination of the Taylor mass powers of the nodes, that is of the 1PI vertices. Consider then a generic vertex function 
\beq
\Gamma^{(r+2s)}_{A^r(c\bar c)^s}\,,\nonumber
\eeq
with $\smash{r+2s\geq 2}$,\footnote{We assume that the original graph is connecete and 1PI. The nodes have then at least two legs.} and let us denote its Taylor mass power by $\omega_{rs}$. If no symmetries were present, the Taylor mass power would equal the mass dimension of the vertex, that is $\smash{4-r-2s}$. The derivative nature of the ghost-antighost-gluon vertex, which is directly related to the anti-ghost shift symmetry of the model (\ref{eq:cf-model}), adds a factor $P$ for each anti-ghost leg. The same derivative nature of the vertex combined with the transverse nature of the gluon propagator (\ref{eq:prop}) adds also a factor $P$ for each ghost leg. This is because this ghost leg of momentum $k$ connects to an internal propagator $G_{\mu\nu}(q)$ of the node, thus producing a factor $\smash{(k+q)_\mu P^\perp_{\mu\nu}(q)=k_\mu P^\perp_{\mu\nu}(q)}$ which vanishes as $k\to 0$.\footnote{Here, it is crucial that the gluon propagators inside the nodes of an AI subgraph reconstruct the full transverse propagator (\ref{eq:prop}), as we have explained above.} Finally, in the case where $r$ is odd, Lorentz symmetry adds an extra factor of $P$. We then arrive at 
\beq
\omega_{rs}=4-r-4s\,,\label{eq:wmn1}
\eeq
for $r$ even, and 
\beq
\omega_{rs}=3-r-4s\,,\label{eq:wmn2}
\eeq
for $r$ odd. 

We note that, for $s\geq 1$, the previous counting applies only to loop corrections  a priori since it requires the presence of internal gluon propagators attached to the ghost external legs. This is fine, however, because most of the vertex functions with external ghost legs do not have a tree-level counterpart. The only two exceptions are $\smash{(r,s)=(0,1)}$ and $\smash{(r,s)=(1,1)}$. In the first case, the formula $\smash{\omega_{01}=0}$ is also valid for the tree-level term $p^2$ which anyway does not appear as a node to an AI subgraph. On the contrary, in the second case, the formula $\smash{\omega_{11}=-2}$ does not apply to the tree-level term since the latter as a 
Taylor mass power equal to $0$. We note, however, that a tree-level ghost-antighost-gluon vertex can only appear as a node at the edge of a tree and we can thus decide not to include it in the AI subgraph. This means that, without loss of generality, we can assume $\smash{\omega_{11}=-2}$.

\begin{figure}[t]
\begin{center}
\includegraphics[width=0.45\textwidth]{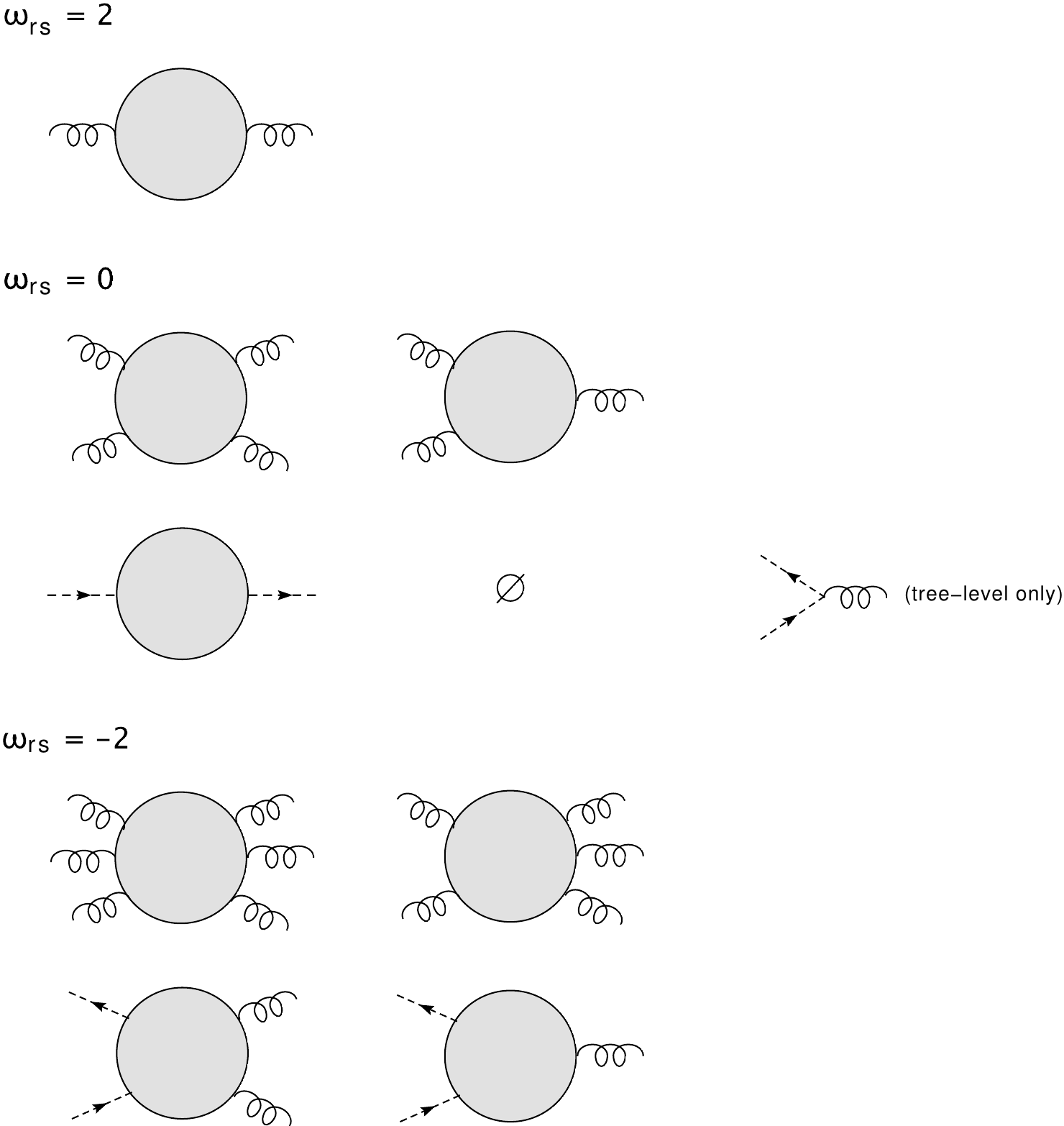}
\caption{Classification of the vertex functions according to their Taylor mass power. The ghost-antighost-gluon vertex with $\smash{\omega=-2}$ refers to the loop corrections only.}
\label{fig:classification}
\end{center}
\end{figure}

From the above formulas, and for later purpose, it is convenient to classify the various vertex functions according to their decreasing Taylor mass power. As already mentionned, we restrict to original graphs that are connected and 1PI and then the nodes that can appear in the AI subgraphs are such that $\smash{r+2s\geq 2}$. Combining this constraint with (\ref{eq:wmn1}) or (\ref{eq:wmn2}), it is easily seen that the highest Taylor mass power is $2$ corresponding to $\Gamma^{(2)}_{AA}$. The next possible Taylor mass power is $0$, corresponding either to $\Gamma^{(4)}_{AAAA}$, $\Gamma^{(3)}_{AAA}$, $\Gamma^{(2)}_{c\bar c}$ and $\Gamma^{(3){\rm tree}}_{Ac\bar c}$.\footnote{As already mentioned, the latter does enter as a node to the AI subgraphs, by definition.} We then have the Taylor mass power $-2$ corresponding to $\Gamma^{(6)}_{A^6}$, $\Gamma^{(5)}_{A^5}$, $\Gamma^{(4)}_{A^2c\bar c}$ and $\Gamma^{(3){\rm loops}}_{Ac\bar c}$. 

The generic structure is quite simple. Corresponding to the Taylor mass power $\smash{\omega=-2k}$, with $k\geq 1$, we will have $\Gamma^{(4+2k)}_{A^{4+2k}}$ and $\Gamma^{(3+2k)}_{A^{3+2k}}$. We will also have all other vertex functions obtained from these two by replacing tetrads of gluon legs by two pairs of ghost-antighost legs. In the case where $k$ is odd, there is the same number of functions derived from $\Gamma^{(4+2k)}_{A^{4+2k}}$ and $\Gamma^{(3+2k)}_{A^{3+2k}}$, whereas in the case where $k$ is even (and thus $4+2k$ a multiple of $4$), there will be one extra function generated from $\Gamma^{(4+2k)}_{A^{4+2k}}$, namely $\Gamma^{(2+k)}_{(c\bar c)^{1+k/2}}$. The classification of the various vertex functions according to their Taylor mass power is represented in Fig.~\ref{fig:classification}.

\subsection{Asymptotic mass powers in the CF model}\label{sec:asym}
We are finally ready for elucidating the origin of the various asymptotic mass powers of a given vertex function. We could analyze each vertex function one after the other. However, we shall proceed in a more efficient way. First, we will show that there is a maximal asymptotic mass power that can be reached among all vertex functions. We will also show that it can only occur in one particular vertex function and that it has a very specific origin. We shall next consider the next-to-maximal asymptotic mass power, show that it occurs only in a restricted class of vertex functions and than it involves again very specific structures. We will do the same for the next-to-next-to-maximal asymptotic mass power before unveiling the all-order structure.

\begin{figure}[t]
\begin{center}
\includegraphics[width=0.14\textwidth]{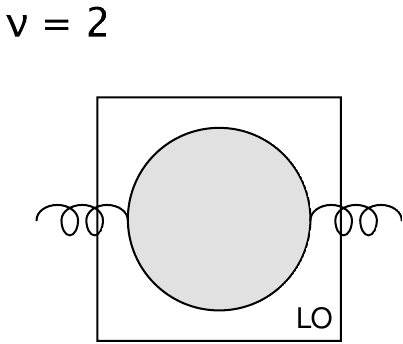}
\caption{Origin of the asymptotic mass power $\smash{\nu=2}$. The box surrounding the AI subgraph represents the leading order (LO) Taylor expansion of the subgraph with respect to its external momenta.}
\label{fig:one}
\end{center}
\end{figure}

\subsubsection{Highest asymptotic mass power}
Let us first show that the asymptotic mass power of any term in any vertex function is at most $2$ and that such an asymptotic mass power occurs in only one very specific case. According to Eq.~(\ref{eq:w}) and because the nodes have at most a Taylor mass power equal to $2$, with $2$ corresponding to $\Gamma^{(2)}_{AA}$, the only way to create a positive mass power is that some of the nodes are gluon self-energy insertions. However, except in the case where the original graph is a contribution to the gluon two-point function, any gluon self-energy insertion will come with two gluon lines. These could be branches of the tree forming the AI subgraph (corresponding to massive components of the gluon propagator) or gluon lines left out of the AI subgraph (because they correspond to massless components of the gluon propagator). 

In any case, each of these gluon lines contributes as $-2$ to the Taylor mass power, thus decreasing effectively the Taylor mass power of the gluon self-energy insertion from $2$ to $-2$. The same is true for chains of gluon self-energy insertions: when including the propagators that connect them, they contribute effectively as $-2$. It follows that the only case where a strictly positive asymptotic mass power appears, $\smash{\nu=2}$, is that of the gluon two-point function, more precisely in the case where the considered AI subgraph is the two-point function itself, Taylor-expanded to leading order. This is illustrated in Fig.~\ref{fig:one}. This asymptotic mass power is necessarily the largest one present in the two-point function and thus corresponds to the (leading) asymptotic mass power of the two-point function.

\subsubsection{Next-to-highest asymptotic mass power}
Let us next investigate how a vanishing asymptotic mass power can occur. Obviously, it could occur in the same case that produces the asymptotic mass power equal to $2$, that is the gluon two-point function with an AI subgraph equal to the graph itself, in the case where the AI subgraph is Taylor-expanded to next-to-leading order. 

If we leave this trivial case aside, it is interesting to remark that the gluon self-energy nodes of the AI subgraph always appear in chains that correspond to dressed gluon lines contributing as $-2$ to the Taylor mass power. We can thus eliminate from Eq.~(\ref{eq:w}) the gluon self-energy nodes and redefine $I_A$ as counting the gluon chains connecting these nodes. It is important to stress that there could remain self-energy insertions in the rest of the graph, made of gluon lines corresponding to massless componnents. For this reason, we should also count in $I_A$ the trivial chains made of one line not connected to any gluon self-energy node. With Eq.~(\ref{eq:w}) modified in this way, all terms are $\leq 0$ and it becomes clear that the only possibility for creating a vanishing asymptotic mass power is, without making use of any of these gluon chains, to combine the various vertex functions with a vanishing Taylor mass power, namely\footnote{Recall that we excluded $\Gamma^{(3){\rm tree}}_{Ac\bar c}$ from the possible nodes, see above.} $\Gamma^{(4)}_{AAAA}$, $\Gamma^{(3)}_{AAA}$, $\Gamma^{(2)}_{c\bar c}$.

\begin{figure}[t]
\begin{center}
\includegraphics[width=0.32\textwidth]{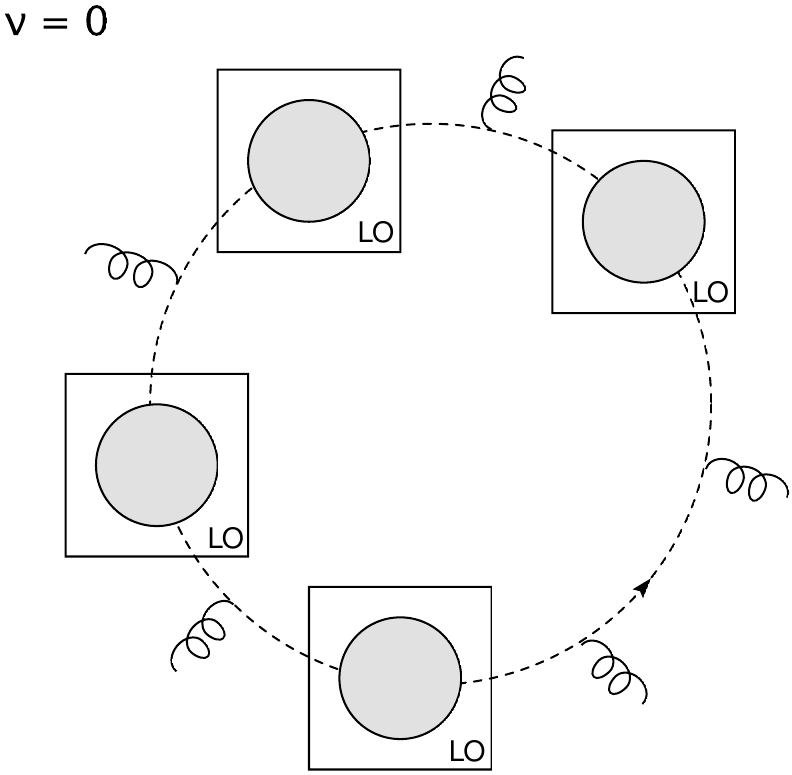}
\caption{Origin of the asymptotic mass power $\smash{\nu=0}$. The boxes surrounding subgraphs represent the leading order (LO) Taylor expansion of the corresponding subgraph with respect to its external momenta. We have not represented the trivial cases leading to $\smash{\nu=0}$, see the main text.} 
\label{fig:two}
\end{center}
\end{figure}

There are again some trivial cases that can be considered, those where the original graph is one of the above functions and the AI subgraph coincides with the graph itself, Taylor-expanded to leading order. If we want non-trivial AI subgraphs that combine various of these functions, we are very limited since no gluon chain is allowed. This means that the trees of the AI subgraph are single nodes and moreover, these cannot be purely gluonic nodes. We are then left with ghost self-energy nodes  $\Gamma^{(2)}_{c\bar c}$ that need to be connected to each other using ghost lines or tree-level ghost-antighost-gluon vertices (that do not belong to the AI subgraph) in order to reconstruct the original graph. Since the latter is assumed to be connected and 1PI, it is clear that the only possibility is to form a single ghost loop connecting the ghost self-energies and as many tree-level ghost-antighost-gluon vertices as wanted. We have then found that, aside from the trivial cases, the only structure generating a vanishing asymptotic mass power is an effective one ghost loop contributing to $\Gamma^{(r\geq 2)}_{A^r}$. This structure is shown in Fig.~\ref{fig:two}.

In summary, the only vertex functions with a vanishing (leading) asymptotic mass power are $\Gamma^{(3){\rm tree}}_{Ac\bar c}$, $\Gamma^{(2)}_{c\bar c}$, $\Gamma^{(r\geq 3)}_{A^r}$. The origin of the vanishing asymptotic mass power is either the AI subgraph coinciding with the graph itself, in the case of $\Gamma^{(3){\rm tree}}_{Ac\bar c}$,  $\Gamma^{(2)}_{c\bar c}$, $\Gamma^{(3)}_{A^3}$ and $\Gamma^{(4)}_{A^4}$, or, in the case of $\Gamma^{(r\geq 3)}_{A^r}$, an effective loop connecting tree-level ghost-antighost-gluon vertices via chains of ghost self-energy insertions, Taylor-expanded to leading order. We stress that this structure also contributes a vanishing asymptotic mass power to $\Gamma^{(2)}_{A^2}$ but it is not the leading asymptotic mass power of that vertex, as we saw in the previous section.

\subsubsection{Next-to-next-to-highest asymptotic mass power}\label{sec:NNLO}
To start unveiling the general structure, let us now consider the case $\smash{\nu=-2}$. Again, one source for such asymptotic mass power is the gluon two-point function, in the case where the AI subgraph is the graph itself, Taylor-expanded to next-to-next-to-leading order. Leaving this case aside, we can again use Eq.~(\ref{eq:w}) not counting the gluon self-energy nodes while counting in $I_A$ only the gluon chains that connect these nodes (and including the trivial chains). In this case, all the nodes contribute negatively to the mass power, which implies that $\smash{I_A=0}$ or $\smash{I_A=1}$. 

If we take $\smash{I_A=0}$, we have of course all the previous structures with AI subgraphs expanded to next-order. In particular, we have the structure of Fig.~\ref{fig:two} with one of the ghost self-energies Taylor-expanded to next-to-leading order. But we can also include new vertices as nodes of the AI subgraph, with Taylor mass power equal to $-2$. There can be only one such vertex. Basically, it dresses one of the ghost-antighost-gluon vertices in the previously obtained effective one ghost loop contributions to $\Gamma^{(m)}_{A^m}$. But we can also form a similar loop where one of the vertices is $\Gamma^{(4)}_{A^2c\bar c}$, see Fig.~\ref{fig:three} (top). There is also the trivial case where the graph is one of the vertices with $\smash{\omega=-2}$, see Fig.~\ref{fig:classification} and the considered AI subgraph is the graph itself, Taylor expanded to leading order.

In the case $\smash{I_A=1}$, one can form either an effective two ghost loop contribution to $\Gamma^{(r)}_{A^r}$, or an effective one ghost loop closed by this gluon chain contributing to $\Gamma^{(r)}_{A^rc\bar c}$. There is also a tree-level contribution but it should not be considered for it is 1PR. These different possibilities are shown in Fig.~\ref{fig:three} (bottom). In particular, since $\Gamma^{(r)}_{A^rc\bar c}$ is the only new class of vertex functions that has appeared, and with the exception of $\Gamma^{(3){\rm tree}}_{A c\bar c}$ and $\Gamma^{(2)}_{c\bar c}$ which already appeared in the analysis of $\smash{\nu=0}$, these are the only vertex functions whose (leading) asymptotic mass power equals $-2$. This leading mass power comes again from an effective one-loop contribution, and also, in the particular case of $\Gamma^{(3){\rm loops}}_{A c\bar c}$ and $\Gamma^{(4)}_{A^2c\bar c}$, from an AI subgraph that coincides with the graph itself.

Let us finally mention that the two last structures of Fig.~\ref{fig:three} involve one gluon line which represents in fact a chain of self-energy insertions. While the latter are necessarily part of the AI subgraph, the lines connecting them should be included in the AI subgraph if they are massive and left out if they are massless. This needs to be taken into account in practice when evaluating the various contributions to $\smash{\nu=-2}$.

\begin{widetext}
\begin{center}
\begin{figure}[t]
\includegraphics[width=0.7\textwidth]{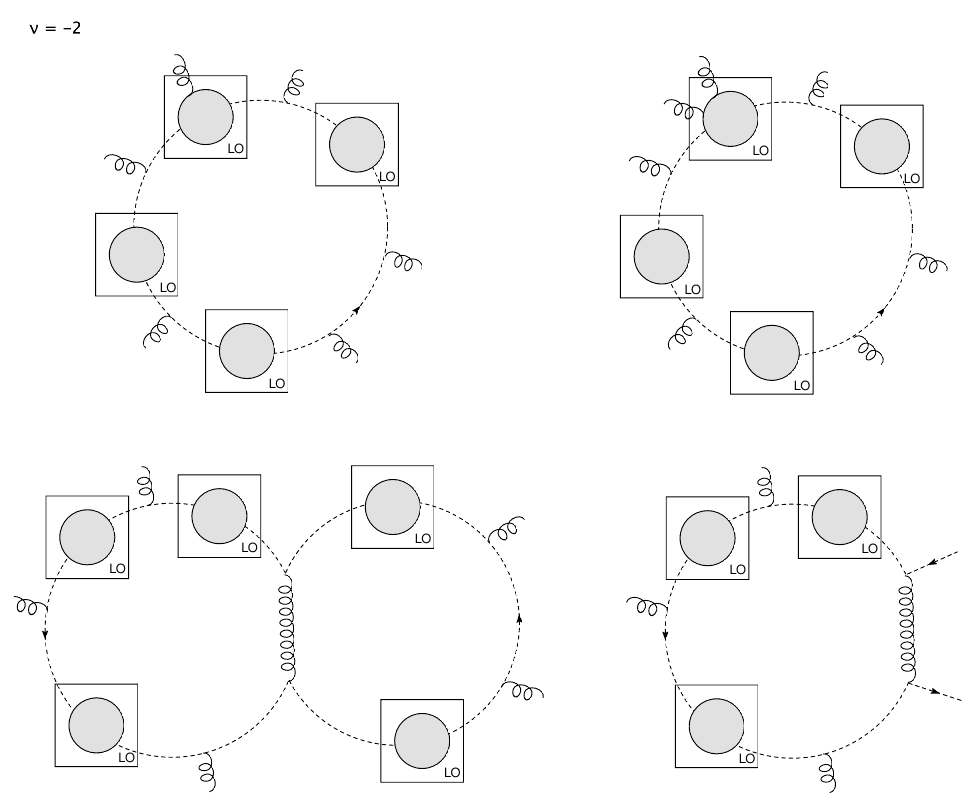}
\caption{Some of the structures contributing to the asymptotic mass power $\smash{\nu=-2}$ (the whole list is discussed in the main text). The boxes surrounding subgraphs represent the leading order (LO) Taylor expansion of the corresponding subgraph with respect to its external momenta. The gluon lines represent chains of gluon self-energy insertions.}
\label{fig:three}
\end{figure}
\end{center}
\end{widetext}

\subsubsection{All-order asymptotic mass powers}
Consider an AI subgraph $\bar {\cal G}$ of a graph ${\cal G}$. Suppose that we shrink any node of the subgraph to a tree-level vertex. The graph ${\cal G}$ is transformed into an effective graph involving effective tree-level vertices with $r$ gluon legs and $s$ pairs of ghost-antighost legs. Those effective tree-level vertices that have no counterpart in (\ref{eq:cf-model}) originate necessarily from the shrinking of the nodes of the AI subgraph (we include here two-point vertices) while those which have a counterpart in (\ref{eq:cf-model}) can also be present outside of the AI subgraph. We recall, however, that in the case of ghost-antighost-gluon tree-level effective vertices, those that originate from the shrinking of the nodes of the AI subgraph are necessarily associated to loop corrections of ghost-antighost-gluon vertex. The number of loops of the effective graph is
\beq
L=I_A+I_c-\sum_{r+2s\geq 2} V_{rs}+1\,,\label{eq:L}
\eeq
where $I_A$ and $I_c$ are, respectively, the number of gluon and ghost lines letting aside those hidden in the nodes, and $V_{rs}$ the total number of effective tree-level vertices with $r$ gluon legs and $s$ pairs of ghost-antighost legs. Denoting by $E_A$ and $E_c$, the number of external gluon and ghost legs, we have as usual
\begin{eqnarray}
E_A & = & -2I_A+\sum_{r+2s\geq 2} rV_{rs}\,,\label{eq:EA}\\
E_c & = & -2I_c+\sum_{r+2s\geq 2} 2sV_{rs}\,.\label{eq:Ec}
\end{eqnarray}
Finally, the asymptotic mass power the AI subgraph contributes to  is
\beq
\nu=-2I_A+\sum_{r+2s\geq 2}\omega_{rs}V^*_{rs}\,,\label{eq:wp}
\eeq
where $V^*_{rs}$ counts only the effective vertices that originate in the shrinking of the AI subgraph.\footnote{Since none of the original tree-level vertices of the model contribute to the Taylor mass power, and because $\smash{V^*_{20}=V_{20}}$ and $\smash{V^*_{01}=V_{01}}$, it could seem that it is possible to replace $V^*_{rs}$ by $V_{rs}$ in the previous formula. However, one should pay attention to the fact that the nodes with $\smash{r=s=1}$ count as $\smash{\omega_{11}=-2}$ rather than $0$ as does the tree-level ghost-antighost-gluon vertex outside the AI subgraph.}

If we leave the case $\smash{\nu=2}$ aside (since we have already treated it above) and because $\smash{\omega_{01}=0}$, we can assume $\smash{r+2s\geq 3}$ in Eq.~(\ref{eq:wp}) and count in $I_A$ only the gluon chains connecting gluon self-energy nodes (see the discussion above). We can also see this by noting that the formula rewrites
\beq
\nu=-2(I_A-V^*_{20})+\sum_{r+2s\geq 3}\omega_{rs}V^*_{rs}\,,
\eeq
with $I_A-V^*_{20}$ counting the number of gluon chains, which will be redefined as $I_A$ in what follows. A similar remark applies to Eq.~(\ref{eq:EA}) upon using $\smash{V^*_{20}=V_{20}}$. 

Similarly, the ghost self-energy nodes can be ignored provided one uses $I_c$ to count the ghost chains that connect these nodes. Again this can be seen from the fact that the terms with $\smash{r+2s=2}$ in Eqs.~(\ref{eq:L}) and (\ref{eq:Ec}) can be absorbed into the redefinitions $\smash{I_A-V_{20}\to I_A}$ and $\smash{I_c-V_{01}\to I_c}$. In what follows, we shall thus work with the set of equations
\begin{eqnarray}
L & = & I_A+I_c-\sum_{r+2s\geq 3} V_{rs}+1\,,\label{eq:19}\\
E_A & = & -2I_A+\sum_{r+2s\geq 3} rV_{rs}\,,\\
E_c & = & -2I_c+\sum_{r+2s\geq 3} 2sV_{rs}\,\label{eq:21}\\
\nu & = & -2I_A+\sum_{r+2s\geq 3} \omega_{rs}V^*_{rs}\,.\label{eq:22}
\end{eqnarray}
If we multiply the last equation by $-1$, we obtain a very interesting result:
\beq
-\nu=2I_A+\sum_{r+2s\geq 3} (-\omega_{rs})V^*_{rs}\,,
\eeq
Since we excluded the case $\smash{\nu=2}$, both the LHS and each of the terms in the RHS are positive. It follows in particular that $I_A$ is bounded:
\beq
I_A\leq -\nu/2\,.
\eeq
which we have verified in the examples above.\footnote{Similarly for those $rs$ such that $V^*_{rs}\neq 0$, we deduce that $\omega_{rs}\geq\nu$, while for those $rs$ such that $\omega_{rs}\neq 0$, we find $V^*_{rs}\leq \nu/\omega_{rs}$.}

\begin{figure}[t]
\begin{center}
\includegraphics[width=0.4\textwidth]{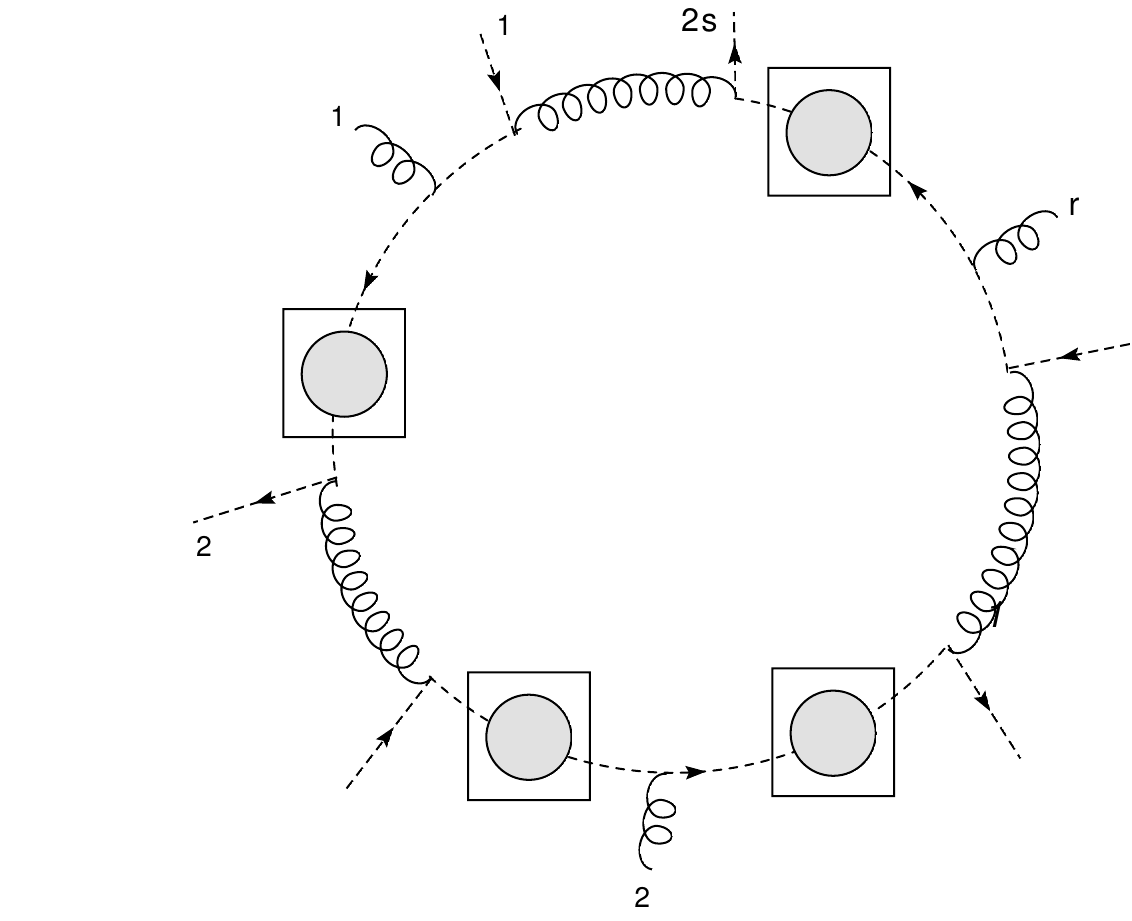}
\caption{Dominant infrared contribution to $\Gamma^{(r+2s)}_{A^r(c\bar c)^s}$. The gluon lines need to be interpreted as gluon chains, see the main text, connecting gluon self-energy nodes. The connecting gluon lines can correspond either to massive components or massless components of the gluon propagator but the latter should not be included in the AI subgraph (for this type of topology).}
\label{fig:one_loop}
\end{center}
\end{figure}

Another interesting equation can be obtained by multiplying Eq.~(\ref{eq:19}) by $2$ and adding both Eqs.~(\ref{eq:21}) and (\ref{eq:22}). We find
\beq
2L+E_c+\nu=\!\!\!\!\sum_{r+2s\geq 3} (2s-2)V_{rs}+\!\!\!\!\!\!\sum_{r+2s\geq 3} \omega_{rs}V^*_{rs}+2\,.
\eeq
It is now interesting to note that for those effective tree-level vertices that have no counterpart in the original model, $\smash{V^*_{rs}=V_{rs}}$. For the others, corresponding to $\smash{(r,s)=(3,0)}$, $(4,0)$ and $(1,1)$, in general we have $\smash{V^*_{rs}\neq V_{rs}}$ but we note that $\smash{\omega_{rs}=0}$ and $\smash{2s-2=-2}$ for the first two, and reversely, $\smash{\omega_{rs}=-2}$ and $\smash{2s-2=0}$ for the last one. We can thus rewrite the previous equation as\beq
2-\nu=2L+E_c+2(V_3+V_4+V^*_{11})+\!\!\!\!{\sum_{r+2s\geq 3}}^{\!\!\!\!\!\!\!*} (2-\omega_{rs}-2s)V_{rs}\,,\label{eq:rhs}
\eeq
where the $*$ on the summation symbol means that we consider only the effective tree-level vertices that have no counterpart in the original model and $V^*_{11}$ counts only the effective tree-level ghost-antighost-gluon vertices that originate from nodes of the AI subgraph. We observe that the tree-level ghost-antighost-gluon vertices that are left outside the AI subgraphs are not at all constrained by this formula.

The coefficients in front of $L$, $E_c$, $V_3$, $V_4$, $V^*_{11}$ in Eq.~(\ref{eq:rhs}) are all strictly positive. The same is true for the prefactor $\smash{2-\omega_{rs}-2s}$ as it can be easily checked. This implies that, for a given $\nu$, the number of effective loops as well as the number of ghost-antighost leg pairs are bounded from above. The same is true  for the number of effective tree-level vertices of each type, with the important exception of the number of tree-level ghost-antighost-gluon vertices that can appear outside the AI subgraphs. 

We can now use these ideas to identify the vertex functions whose (leading) asymptotic mass power is $\smash{\nu\leq 0}$ while unveiling the particular structures that contribute to it. To see how this works, let us first see how the results of Sec.~\ref{sec:NNLO} are retrieved. We saw before that the vertex functions whose (leading) asymptotic mass power is $\smash{\nu\geq 0}$ are all the vertex functions including only gluons and also the ghost two-point vertex. This means that, if we want to find vertex functions with $\smash{\nu=-2}$, we need to add one pair of ghost-antighost legs, that is $\smash{E_c\geq 2}$. On the other hand, from (\ref{eq:rhs}), we have $\smash{E_c\leq 4}$. Still from (\ref{eq:rhs}), we see that choosing $\smash{E_c=4}$ saturates the bound and imposes $\smash{L=0}$ as well as $\smash{V_3=V_4=V^*_{11}=0}$, and similarly $\smash{V_{rs}=0}$ for all effective tree-level vertices without counterpart in the original model. This case is clearly inconsistent. We are then left with the choice $\smash{E_c=2}$ which from (\ref{eq:rhs})  leaves open the two possibilities $\smash{L=0}$ or $\smash{L=1}$. In the first case, the AI subgraph should coincide with the graph itself and, therefore, the vertex functions should be chosen among those with $\smash{\omega_{rs}=-2}$ and $\smash{s=1}$ (since $\smash{E_c=2}$), see Fig.~\ref{fig:classification}. We are then left with $\Gamma^{(3){\rm loops}}_{Ac\bar c}$ and $\Gamma^{(4)}_{A^2c\bar c}$, as we already saw in (\ref{sec:NNLO}). In the second case, $\smash{L=1}$, the bound is saturated already with $2L+E_c$, the only possible vertices available are the tree-level ghost-antighost-gluon vertices and the structure is the one corresponding to the last diagram of Fig.~\ref{fig:three}. 

To continue the recursion, the key point is that all vertex functions with at most one pair of ghost-antighost legs and an arbitrary number of gluons legs have been attributed a (leading) asympotic mass power $\smash{\nu\geq -2}$. This means that, if one now looks for vertex functions with $\smash{\nu=-4}$, one needs to consider $\smash{4\leq E_c\leq 6}$, with only the case $\smash{E_c=4}$ leading to sensible results. Because of the bound, we have again $\smash{L=0}$ or $\smash{L=1}$. The first case, requires us to choose the vertex functions among those with $\smash{\omega_{rs}=-4}$ and $\smash{s=2}$, which leaves only the possibility $\Gamma^{(4)}_{(c\bar c)^2}$. In the second case, the only possible vertices available are again the tree-level ghost-antighost-gluon vertices and the structure is the one corresponding to Fig.~\ref{fig:one_loop} with $\smash{s=2}$. After this, all vertex functions with at most two pair of ghost-antighost legs and an arbitrary number of gluons legs have been attributed a (leading) asympotic mass power $\smash{\nu\geq -4}$ and the recursion can start over again.

At the end of the day, we obtain that the vertex functions with (leading) asymptotic mass power $\nu\leq 0$ are\footnote{The only exception to this rule is $\Gamma^{(2)}_{c\bar c}$ whose leading asymptotic mass power is $\smash{\nu=0}$.}  $\Gamma^{r-\nu}_{A^r(c\bar c)^{-\nu/2}}$ and the structure leading to this behavior is that of Fig.~\ref{fig:one_loop} with $s=-\nu/2$ or the AI subgraph coinciding with the graph itself. In this later case, since $\nu=\omega_{r,-\nu/2}$ one finds that $r=4+\nu$ or $r=3+\nu$ depending of whether $r$ is even or odd. In particular, this structure does not contribute to $\nu$ as soon as $\nu\leq -6$ and only the one in Fig.~\ref{fig:one_loop} matters in this case.

\subsection{Structure of leading logarithms}\label{sec:logs}
In this section, we restrict to particular configurations of the external momenta, such as the one studied in the main text, that depend on one scale only. By this, we mean that $\smash{p_i=p\,a_i u_i}$, with $u_i$ unit vectors, $a_i\geq 0$ and $p>0$. We want to discuss the presence of logarithms in the leading asymptotic expansion as $p\to 0$.\footnote{We are implicitly assuming that these configurations do not lead to IR divergences in the loops for $p\neq 0$ even though divergences could appear in the limit $p\to 0$.}

We have seen above that for each vertex function $\Gamma^{(r+2s)}_{A^r(c\bar c)^s}$, with $r+2s\geq 3$, the leading asymptotic mass power in a large mass expansion is $\nu=-2s$ and arises from the one-loop effective structure represented in Fig.~\ref{fig:one_loop}, and, in some cases, also from AI subgraphs that coincide with the graph itself. Since the latter have a regular dependence on $p$ as $p\to 0$, and since we are after logarithms here, we focus on the diagrams of Fig.~\ref{fig:one_loop}. 

These diagrams contribute as $m^{-2s+c\epsilon}p^{r+2s+d\epsilon}$, for some $c$ and $d$, times various integrals, one being the explicit loop in the diagram and the others being associated with the leading coefficient of the Taylor-expanded ghost self-energy insertions. Since we have factored out the whole $m$ and $p$ dependences, the integrals that arise from the self-energies depend only on $\epsilon$, while the explicit one-loop integral may depend on the $a_i u_i$. When expanded in $\epsilon$, these integrals may contain poles in $1/\epsilon$. However, after renormalization of the self-energy insertions, these poles can only originate in the explicit one-loop integral, in the case of primitively divergent vertices, which produce a simple pole in $1/\epsilon$.

 From this, we conclude that the leading behavior of any primitively divergent vertex function with $r+2s\geq 3$ has at most linear logarithms in $p$,\footnote{The IR expansion (\ref{eq:gamma-ir-exp}) is compatible with these expectations.} to all orders of perturbation theory, and that no such logarithms are present in the other functions.\footnote{We put here no prejudice on whether the coefficient of the logarithm depends on the $a_iu_i$. This can be very simply investigated.} A similar linear logarithm can be identified to all orders at next-to-leading order in the mass expansion of $\Gamma^{(2)}_{A^2}$ or $\Gamma^{(2)}_{c\bar c}$ since the structure contributing to the next-to-leading asymptotic mass power is again that of Fig.~\ref{fig:one_loop}. On the other hand, nothing prevents the powers of the logarithms in $m$ to grow with the number of loops. For instance, the renormalized self-energy insertions may contain any power of those This means that in order to ensure perturbative control in the IR, one should choose a scale $\mu(p)$ such that $\smash{\mu(p)\to m}$ as $p\to 0$, rather than $\smash{\mu(p)\sim p}$ as we should do in the UV. This motivates our choice of scale in the main text.

 \section{One-loop ghost diagram}\label{app:one}
Let us consider the one-loop ghost diagram in the case where the external momenta depend only on one scale $p>0$, that is $\smash{p_i=p a_i u_i}$ with $a_i>0$ and $u_i$ a unit vector. There are two such diagrams (depending on the orientation of the ghost loop) which, up to trivial factors and a color structure $f^{abc}$, lead to the contribution
\beq
\int_q \frac{q_\rho (q+p_1)_\mu(q-p_3)_\nu}{q^2(q+p_1)^2(q-p_3)^2}-\int_q \frac{q_\rho (q+p_2)_\nu(q-p_3)_\mu}{q^2(q+p_2)^2(q-p_3)^2}\,.
\eeq
Rescaling $q$ as $pq$ and renaming $p_i$ as $p_i=a_iu_i$, we can factor out the complete $p$ dependence as
\beq
p^{1-2\epsilon}\left[\int_q \frac{q_\rho (q+p_1)_\mu(q-p_3)_\nu}{q^2(q+p_1)^2(q-p_3)^2}-\int_q \frac{q_\rho (q+p_2)_\nu(q-p_3)_\mu}{q^2(q+p_2)^2(q-p_3)^2}\right],
\eeq
where both $q$ and the $p_i$ are dimensionless in this last formula. The origin of the logarithm in $p$ is then pretty clear. Upon expansion in $\epsilon$, there will be a term $\epsilon \ln p$ multiplying the pole in $1/\epsilon$ stemming from the integrals. The latter can be obtained by expanding the integrand at large $q$ and can be seen to correspond to the $1/\epsilon$ pole of the integral $A\equiv\int_q 1/(q^2+m^2)^2$ times the tensor
\begin{eqnarray}\label{eq:D3}
& & \frac{1}{d}\left[(p_{1\mu}+p_{3\mu})\delta_{\nu\rho}-(p_{2\nu}+p_{3\nu})\delta_{\mu\rho}\right]\nonumber\\
& & -\,\frac{2}{d(d+2)}(p_1-p_2)_\sigma[\delta_{\mu\nu}\delta_{\rho\sigma}+\delta_{\mu\rho}\delta_{\nu\sigma}+\delta_{\mu\sigma}\delta_{\nu\rho}],\nonumber\\
\end{eqnarray}
where we have used 
\begin{eqnarray}
\int_q q_\mu q_\nu f(q^2) & = & \frac{\delta_{\mu\nu}}{d} \int_q q^2 f(q^2)\,,\nonumber\\
\int_q q_\mu q_\nu q_\rho q_\sigma f(q^2) & = & \frac{\delta_{\mu\nu}\delta_{\rho\sigma}+\delta_{\mu\rho}\delta_{\nu\sigma}+\delta_{\mu\sigma}\delta_{\nu\rho}}{d(d+2)} \int_q q^4 f(q^2)\,.\nonumber
\end{eqnarray} Since the pole is already contained in the integral $A$, we can set $d=4$ in Eq.~(\ref{eq:D3}). Moreover, using that $\smash{p_1+p_2+p_3=0}$, the latter becomes
\beq
-\frac{1}{12}[(p_1-p_2)_\rho\delta_{\mu\nu}+(p_2-p_3)_\mu\delta_{\nu\rho}+(p_3-p_1)_\nu\delta_{\rho\mu}].
\eeq
We conclude that the leading (logarithmic) IR behavior of the one-loop ghost diagram in the considered regime has the same structure than the tree-level tensor component. This extends to the exact three-gluon vertex since the dominant contribution in this regime is that of the one-loop ghost diagram times the cube of the exact ghost dressing function at zero-momentum, see the main text for details.

\bibliographystyle{apsrev4-1}
\bibliography{refs}

\end{document}